\documentclass{sig-alternate}

\newif\ifcomm
\commtrue

\usepackage{cite}
\usepackage{textcomp}
\usepackage{array}
\usepackage[tight,footnotesize]{subfigure}
\usepackage{graphicx}
\usepackage{stfloats}
\usepackage{url}
\usepackage{xcolor}
\usepackage{algorithm}
\usepackage{algorithmic}
\usepackage{todonotes}
\usepackage{appendix}
\newcommand{\ORM}[1]{}

\newcommand{\PR}[1]{#1}

\newcommand{\RM}[1]{#1}
\newcommand{\TR}[1]{#1}
\newcommand{\T}[1]{\noindent\textbf{#1}}
\newcommand{\ST}[1]{\noindent\textbf{#1}.}
\newcommand{\K}[1]{#1}

\newcounter{commentNumberI}
\ifcomm
\newcommand{\I}[1]{\addtocounter{commentNumberI}{1} \todo[inline,color=blue!20]{\textbf{(I.\arabic{commentNumberI})} #1}}
\else
\newcommand{\I}[1]{}
\fi

\newtheorem{theorem}{Theorem}
\newtheorem{property}{Property}
\newtheorem{lemma}{Lemma}
\newtheorem{corollary}{Corollary}

\newcommand{\eq}[1]{Equation~(\ref{#1})}        
\newcommand{\thm}[1]{Theorem~\ref{#1}}
\newcommand{\lem}[1]{Lemma~\ref{#1}}

\newcommand{\bp}{\begin{proof}}
\newcommand{\ep}{\end{proof}}

\newcommand{\be}{\begin{equation}}
\newcommand{\ee}{\end{equation}}

\newcommand{\fig}[1]{Fig.~\ref{#1}}

\newcommand{\para}[1]{\left( #1 \right)}        
\newcommand{\brac}[1]{\left\{ #1 \right\}}
\newcommand{\sbrac}[1]{\left[ #1 \right]}
\newcommand{\floor}[1]{\left \lfloor #1 \right \rfloor}
\newcommand{\ceil}[1]{\left \lceil #1 \right \rceil}

\newcommand{\abs}[1]{\left\vert#1\right\vert}

\begin{document}

\setlength{\pdfpagewidth}{8.5in}
\setlength{\pdfpageheight}{11in}
%
\title{Links as a Service (LaaS):\\Guaranteed Tenant Isolation in the Shared Cloud}
\author{
\alignauthor Eitan Zahavi\footnotemark[1] \footnotemark[4] , Alexander Shpiner\footnotemark[4] , Ori Rottenstreich\footnotemark[5] , Avinoam Kolodny\footnotemark[1]\; and Isaac Keslassy\footnotemark[1] \footnotemark[6]  \\
\affaddr{\footnotemark[1]\; Technion \quad \footnotemark[4]\; Mellanox \quad \footnotemark[5]\; Princeton\quad \footnotemark[6]\; VMware}
}



\maketitle



\maketitle

\begin{abstract}
The most demanding tenants of shared clouds require complete isolation from their neighbors, in order to guarantee that their application performance is not affected by other tenants. Unfortunately, while shared clouds can offer an option whereby tenants obtain dedicated servers, they do not offer any {network provisioning} service, which would shield these tenants from network interference. 

In this paper, we introduce \emph{Links as a Service (LaaS)}, a new abstraction for cloud service that provides isolation of network links. Each tenant gets an exclusive set of links forming a virtual fat-tree, and is guaranteed to receive the exact same bandwidth and delay as if it were alone in the shared cloud. {Consequently, each tenant can use the forwarding method that best fits its application.}
Under simple assumptions, we derive theoretical conditions for enabling LaaS without capacity over-provisioning in fat-trees.
New tenants are only admitted in the network when they can be allocated hosts and links that maintain these conditions.
LaaS is implementable with common network gear, tested to scale to large networks and provides full tenant isolation at the worst cost of a 10\% reduction in the cloud utilization.
\end{abstract}

\sloppy

\section{Introduction}\label{sec_introduction}
Many owners of private data centers would like to move to a shared multi-tenant cloud, which can offer a reduced cost of ownership and better fault-tolerance.
For some of these tenants it is vital that their applications will not be affected by other tenants, and will keep exhibiting the same performance\footnote{By \emph{performance}, we refer to the inverse of either the total application run-time, including both the computation and communication times, or of the response time of online services.}\cite{linden2006make,mmayer2009speed,artz2009secret}.
{For example, a banking application may need to roll-up all accounts data overnight, and a weather prediction software should similarly complete within a highly predictable time. For such tenants, run-time predictability is a key requirement.}

Unfortunately, distributed applications often suffer from unpredictable performance when run on a shared cloud~\cite{ballani_towards_2011,iosup_performance_2011}. This unpredictable performance is mainly caused by two factors: \emph{server sharing} and \emph{network sharing}~\cite{mogul_what_2012,curtis_mahout:_2011,iosup_hpc_cloud_2011,ostermann_performance_2010,lacurts_cicada_2014,chowdhury_vineyard:_2012,jokanovic_effective_2012,doriguzzi_corin_vertigo:_2012,lam_netshare_2012,rodrigues_gatekeeper_2011,webb_topology_2011,guo_secondnet_2010,al-fares_hedera:_2010,shieh_seawall_2010,sherwood_flowvisor:_2009,yu_rethinking_2008,raghavan_cloud_2007}.
The first factor, \emph{server sharing}, is easily addressed
by using bare-metal provisioning of servers, such that each server is allocated to a single tenant~\cite{openstack_ironic_2014}.
However, the second factor, \emph{network sharing}, is much more difficult to address. When network links are shared by several tenants, network contention can significantly worsen the application performance if other tenant applications consume more network resources, e.g. if they simply want to benchmark their network or run a heavy backup~\cite{jokanovic_impact_2010}. This can of course prove even worse when other tenants purposely generate adversarial traffic for DoS or side-channel attacks~\cite{ristenpart_hey_2009}.

As detailed in Section~\ref{sec:related_work}, current solutions either (a) require tenants to provide and adhere to a specific traffic matrix declared in advance, which often proves impractical~\cite{chowdhury_vineyard:_2012,yu_rethinking_2008};
(b) follow the hose model by providing enough throughput for any set of admissible traffic matrices~\cite{duffield_flexible_1999,webb_topology_2011,ballani_towards_2011}, but also significantly reduce the link bandwidth and burst size that can be allocated to each VM;
or (c) attempt to track the current traffic matrix, but cannot guarantee constant performance~\cite{raghavan_cloud_2007,lam_netshare_2012,guo_secondnet_2010,shieh_seawall_2010,rodrigues_gatekeeper_2011}.
In addition, while it is known that tailoring the packet forwarding method to the specific tenant application can increase its performance, none of the current cloud solutions allow multiple forwarding algorithms to co-exist on the same network without impacting performance.
\begin{figure*}[]
\centering
\subfigure[No LaaS: Shared links] {
\includegraphics[width=0.6 \columnwidth]{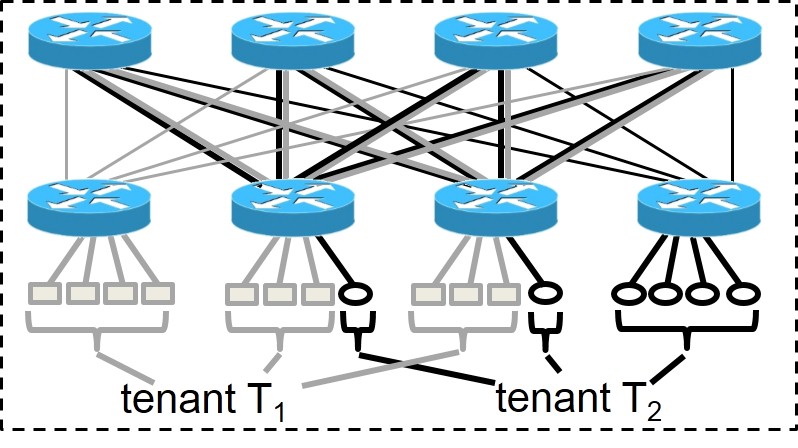}
\label{fig:before_n_after_a}
}
\subfigure[No LaaS: Bandwidth loss] {
\includegraphics[width= 0.6 \columnwidth]{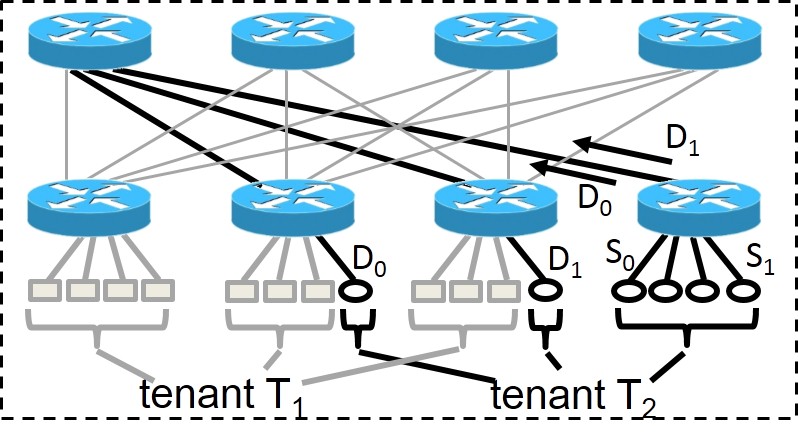}
\label{fig:before_n_after_b}}
\subfigure[LaaS: Full isolation]{ 
\includegraphics[width= 0.6 \columnwidth]{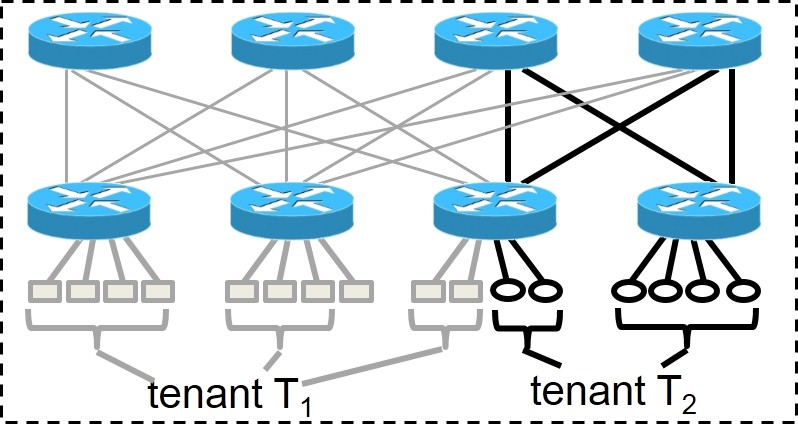}
\label{fig:before_n_after_c}}
\hfill
\caption{Two tenants 
hosted on a cloud. (a) Their traffic interferes on many shared links. (b) There are no shared links, but
the second tenant cannot service an admissible traffic
from $S_0$ and $S_1$ to $D_0$ and $D_1$. (c) Under LaaS conditions of tenant placement and link allocation, the network can service any admissible tenant traffic demands.
}
\label{fig:before_n_after} \vspace{-0.3cm}
\end{figure*}


\emph{In this paper, we introduce a simple and effective approach that eliminates any interference in the cloud network. This approach allows each tenant to use a network forwarding algorithm that is optimized for its own application.}
\K{Keeping with the notion that good fences make good neighbors, we argue that the most demanding tenants should be provided with exclusive access to a subset of the data center links, such that each tenant receives its own dedicated fat-tree network.}
We refer to this cloud architecture model as \emph{Links as a Service (LaaS)}.
The LaaS model guarantees that these tenants can obtain the exact same bandwidth and delay as if they were alone in the shared cloud, independently of the number of additional tenants.
We {show that} allocation of links to tenants is cost-effective and implementable by using common hardware.
{Note that LaaS can similarly support a relaxed model that splits physical links into time-domain-multiplexed channels.
This relaxed model allows multiple tenants per server, but requires accurate packet pacing~\cite{jang_silo_2015} not provided by common hardware today.}

While the LaaS abstraction is attractive, Figure 1 illustrates why it can be a challenge to provide it given any arbitrary set of tenants. First, \fig{fig:before_n_after_a} illustrates a bare-metal allocation of distinct hosts (servers) to two tenants that does not satisfy the LaaS abstraction, since the tenants share common links.
Likewise, the allocation of hosts and links in~\fig{fig:before_n_after_b} also does not satisfy LaaS, even though no links are shared between tenants.
This is because, regardless of the packet forwarding algorithm, internal traffic of the second tenant from the two hosts $S_0$ and $S_1$ in the right leaf switch to hosts $D_0$ and $D_1$ would need to share a common link, and so some admissible traffic patterns would not be able to obtain full bandwidth.
Interestingly, for this host placement, we find that there is in fact no link allocation that can provide full bandwidth to all the admissible traffic patterns of both tenants.
Finally, \fig{fig:before_n_after_c} fully satisfies the LaaS conditions.
All tenants obtain dedicated hosts and links, and can service any admissible traffic demands between their nodes, independently of the traffic of other tenants.
To generalize the above examples, we further analyze the fundamental requirements for providing LaaS guarantees to tenants in 2- and 3-level homogeneous fat-trees. Under minor assumptions, our analysis provides the necessary and sufficient conditions to guarantee the same bandwidth and delay performance over the shared fat-tree networks as when being alone in the shared cloud.
These conditions are novel and greatly reduce the complexity for the online allocation algorithm presented in Section~\ref{sec:analysis}.

We implement a standalone LaaS scheduler that automates tenant placement on top of OpenStack, as well as configures an InfiniBand SDN controller to provide forwarding without interference.
Our open-source code is made available online~\cite{dropbox_laas_2015}.
We show that using this code, our LaaS algorithm responds to tenant requests within a few milliseconds, even on a cloud of 11K nodes, i.e. several orders of magnitude faster than the time it takes to provisioning a new virtual machine. In addition, when the average tenant size is smaller than a quarter of the cloud size, we find that our LaaS algorithm achieves a cloud utilization of about 90\%, for various tenant-size distributions.
For larger tenant sizes, our LaaS allocation converges to the maximal utilization obtained by a bare-metal scheduler that packs tenants without constraints.
{Finally, to demonstrate LaaS strength, we show performance improvements of 50\%-200\% for highly-correlated tenant traffic generated by a Bulk Synchronous Parallel (BSP) application relying on data exchanges along a virtual three-dimensional axis system.}
Thus, the performance improvement exceeds the utilization cost for such applications, uncovering an economic potential (Section~\ref{sec:evaluation}).

While we focus, for brevity, on full-bisectional-bandwidth fat-trees, we show how LaaS can be extended to support over-provisioned (slimmed) fat-trees. We also describe how LaaS can fit more general cloud cases, e.g. when mixing highly-demanding tenants with regular tenants (Section~\ref{sec:discussion}).

\K{Our evaluations show that LaaS is practical and efficient, and completely avoids inter-tenant performance dependence.}

\ORM{
\T{Paper Organization:}
The rest of the paper is structured as follows:
Section~\ref{sec:motivation} provides motivation to our work by showing the impact of cross tenant traffic on application performance.
Section~\ref{sec:architecture} describes the proposed system architecture.
Section~\ref{sec:analysis} presents our analysis of the problem and the tradeoffs in implementation of the LaaS tenant scheduler.
Section~\ref{sec:evaluation} presents evaluation of the proposed system and section~\ref{sec:related_work} describes related work.
} 

\section{Related Work}\label{sec:related_work}

\T{Application variability.} Several studies about the variability of cloud services and HPC application performance were presented by~\cite{iosup_performance_2011,schad_runtime_2010,ballani_towards_2011,jokanovic_impact_2010,orduna_new_2001,bhatele_there_2013}.
They show significant variability for such applications, which strengthens the motivation for using LaaS.

\T{Network isolation.} Specific high-dimensional tori super-computers like IBM BlueGene, Cray XE6, and the Fujitsu K-computer provide scheduling techniques to isolate tenants~\cite{pascual_job_2009,bhatele_there_2013,ajima_tofu:_2009}.
However, they all rely on forming an isolated cube on 3 out of the 5- or 6-dimensional torus space, and thus cannot be used in clouds with fat-tree topologies. They also exhibit a significantly lower cluster utilization, measured as the amount of servers used over time, than the 90\% utilization obtained by LaaS on fat-trees.
Another approach, reduces the interference between jobs running on same fat-tree by applying hard placement constraints~\cite{jokanovic_quiet_2015}. This work reduces but does not guarantee jobs isolation from each other.

\T{Packet forwarding.}
{Many architectures rely on Equal Cost Multiple Path (ECMP)~\cite{choppsmerit_analysis_2015} to spread the allocated tenant traffic and avoid the need to allocate exact bandwidth on each of the used physical links~\cite{ballani_towards_2011,jeyakumar_eyeq_2013,popa_elasticswitch_2013}.
However, while ECMP load-balancing is able to balance the average bandwidth of many small bandwidth flows, it suffers from a heavy tail of the load distribution.
When traffic contains a relatively small number of large flows, ECMP is known to provide poor load-balancing.
Thus, other tenants will affect the application performance. 

Silo~\cite{jang_silo_2015} aims to provide guaranteed latency, bandwidth and burst size to multiple tenants for a worst-case traffic pattern, assuming that tenants do not optimize their forwarding scheme.
Silo achieves its guarantees by applying accurate rate- and burst-size moderation to enforce centrally-calculated values obtained from network calculus.
Unfortunately, Silo does not take forwarding into account. For instance, consider a tenant of 200 VMs placed across more than one 2-level sub-tree (which normally can contain thousands of VMs). If 100 VMs need to send traffic to the other VMs through the same uplink because of the forwarding rules, then each would be restricted to use at most 1/100th of the link bandwidth and 1/100th of the switch buffer size, which is unacceptable for current large tenants. 
LaaS allows the tenants to adapt their forwarding to the traffic pattern without introducing inter-tenant interference, thus allowing them to fully consume the full network bandwidth.

\T{Time separation.} {Some systems like Cicade~\cite{lacurts_cicada_2014} accept the need for handling the varying nature of tenant traffic instead of relying only on the average demand. They assume that traffic demands change at a pace that is slow enough to enable them to react.}
Alternatively, scheduling the MapReduce shuffle stages was proposed by Orchestra~\cite{chowdhury_managing_2011}.
A generalization of this approach that allows a tenant to describe its changing communication needs is suggested by Coflow~\cite{chowdhury_coflow:_2012}.
On the same line of thought, scheduling at a finer grain was proposed by Hedera~\cite{al-fares_hedera:_2010}.
However, since these schemes propose a fair-share network bandwidth to the current set of applications, they actually change the performance of a tenant when new tenants are introduced. Even though fairness does improve, the tenant performance variability grows.

\T{Tenant resource allocation.} Cloud network performance has received significant attention over the last few years.
An overview of the different proposals to allocate tenant network resources is provided by~\cite{mogul_what_2012}.

Virtual Network Embedding maps tenants' requested topologies and traffic matrix over arbitrary clusters~\cite{chowdhury_vineyard:_2012,yu_rethinking_2008}. However, tenants must know and declare their exact traffic demands which is mostly impractical.
Moreover, valid embedding is calculated by variants of linear programming, which are known not to scale as the size of the data centers and number of tenants grow.
In addition, as most of these solutions rely on the tenant traffic matrix, they consider only the average demands,
falling short of representing the dynamic nature of the application traffic.
For example, they prove problematic when an application alternates between several traffic permutations, each utilizing the full link bandwidth.

Other proposals, such as Topology Switching and Oktopus~\cite{webb_topology_2011,ballani_towards_2011}, propose an abstraction for the topology and traffic demands to be allocated to the tenants. They are similar to the hose model proposed for Virtual Private Networks in the context of WAN~\cite{altin_robust_2010}. In addition, \cite{angel_end_to_end_2014} attempts to provide a feedback-based fair-share bandwidth using edge-based rate-limiting. However, to guarantee tenant latency predictability and isolation, such solutions would need strict time-pacing of packets, small limits on allowed VM bandwidth and burst-size allocation, as shown in~\cite{jang_silo_2015}. As mentioned above, these are impractical in current networks.

Another approach for isolation may rely on distributed rate limiting like~\cite{raghavan_cloud_2007}, NetShare~\cite{lam_netshare_2012}, ScondNet~\cite{guo_secondnet_2010}, Seawall~\cite{shieh_seawall_2010}, Gatekeeper~\cite{rodrigues_gatekeeper_2011} and Oktopus~\cite{ballani_towards_2011}.
But distributed rate limiting at the network edge requires tenant-wide coordination to avoid bottlenecks due to load-imbalance.
This coordination leads to response time in the order of milliseconds~\cite{jeyakumar_eyeq_2013}, while the life time of a traffic pattern for high-demanding applications may be 2 to 3 orders of magnitude shorter.

\T{Fairness.} FairCloud provides a generalization of the required fairness properties of the shared cloud network~\cite{popa_faircloud:_2012}.
LaaS tenant isolation satisfies these requirements, and avoids the allocation complexity of the general case.

\T{Application-based routing.} The above schemes for network resource allocation ignore the fact that each tenant application may perform best with a different routing scheme.
Routing algorithm types span a wide range. Some are completely static and optimized for MPI applications~\cite{zahavi_fat-tree_2012,gong_network_2013}.
Others rely on traffic-spreading techniques like ECMP~\cite{choppsmerit_edu_analysis_2000}, rely on traffic spray as in RPS or DeTail~\cite{dixit_impact_2013,zats_detail:_2012}, use adaptive routing as proposed by DARD~\cite{wu_dard:_2012}, or even rely on per-packet synchronized schemes like FastPass~\cite{perry_fastpass:_2014}.
LaaS isolates the sub-topology of each tenant, and therefore allows each tenant to use the routing that maximizes its application performance.
Without link isolation the different routing engines must continuously coordinate the actual bandwidth each one of them utilize from each link.
It is clear that the involved complexity of such scheme renders it slow and impractical.

\section{Impact of Tenant Interference}
\label{sec:motivation}

This section presents the impact of concurrent tenant traffic on tenant performance.
The presented results are obtained from measurements on real hardware, as well as simulations of InfiniBand and Ethernet networks.
We also provide online a full description of the settings 
and of our code for the experiments~\cite{dropbox_laas_2015}.

\T{Tenant interference in cluster experiments.}
\TR{
\begin{figure}[]
\centering
\includegraphics[width= 0.8 \columnwidth]{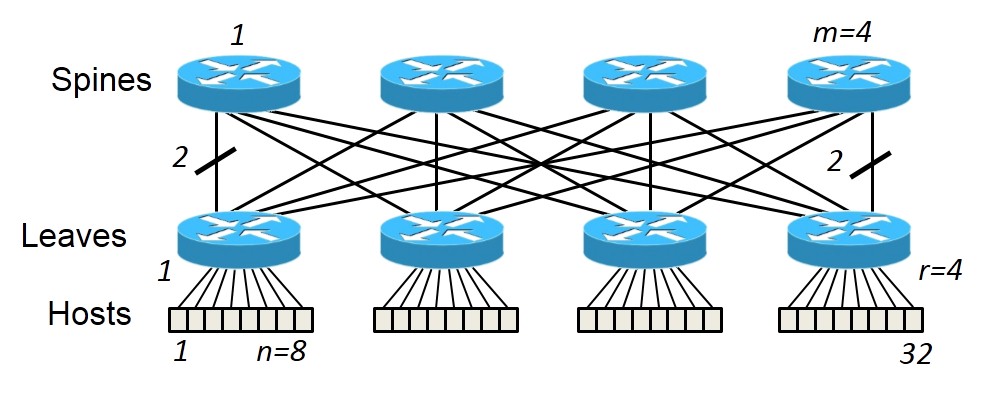}
\caption{Experimental fat-tree cluster.} 
\label{fig:test_cluster}
\end{figure}
}
The experimental topology is a non-blocking two-level fat-tree with 8 hosts in each of the 4 leaf switches.
The leaf switches are fully connected to 4 spine switches, with two parallel links per connection.
We assume 4 tenants, and randomly assign 8 dedicated hosts to each of the 4 tenants.
\RM{The reason for using a random placement is that even a scheduler that follows a bin-packing algorithm is known to show a large degree of fragmentation in steady state~\cite{pascual_job_2009}.}
The tenants independently alternate between computation and all-to-all communication, i.e. each node computes new results and sends different data to the rest of the nodes that belong to the same tenant, as a sequence of un-synchronized shift permutations.
This traffic pattern is representative of the Shuffle stage of MapReduce, and of scientific-computing applications such as those based on Fast Fourier Transform. We keep the total computation time constant, while the communication time changes with the increasing message size (where message means a continuous flow between a pair of machines). For a single tenant with 32KB messages, the communication time represents roughly 2/3 of the total time.

\fig{fig:job_performance_on_cluster} presents the relative application performance in our cluster, measured for various reasonable message sizes~\cite{rabenseifner2001_beff} and for 1--4 parallel tenants.
The results show that even in such a small cluster, the performance of a tenant may degrade (i.e., its run-time may increase) by 25\% for large messages when other tenants run concurrently. Larger message sizes degrade the performance due to the larger buffering needs {and larger communication time.}

Since we also want to analyze the performance of the applications in larger clusters, we further rely on a simulator based on an InfiniBand model~\cite{domke_deadlock-free_2011}. 
For sanity check, we compare our small cluster measurements with simulated results.
The figure illustrates that the simulation results for 4 tenants are about 3\% worse, and show the same trend as the experiment. The difference  probably results from a lack of accuracy in modeling the MPI computation time, and therefore it would be expected to decrease in larger networks with a more significant network contention.
\begin{figure}[t]
\centering
\includegraphics[width= \columnwidth]{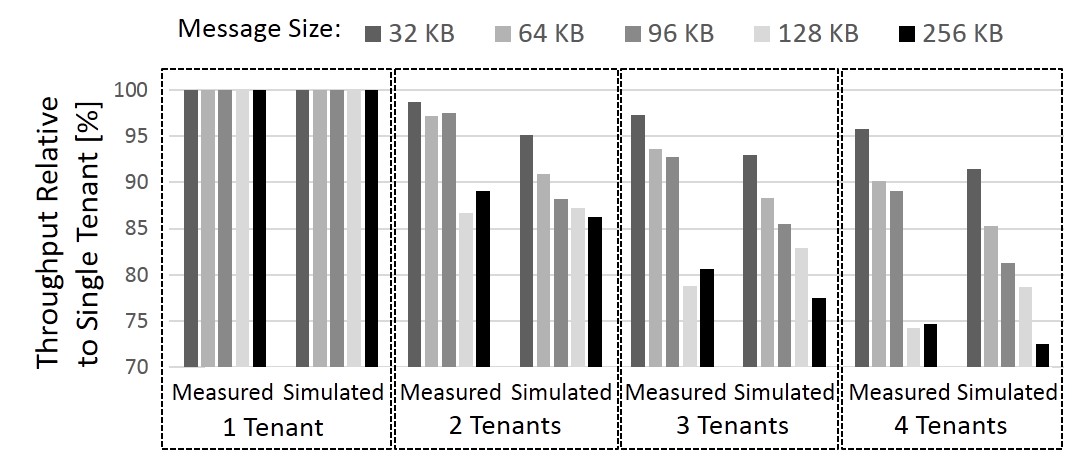}
\caption{Relative performance, obtained by experiment and simulation, of an application based on all-to-all traffic, for 1--4 concurrent tenants of 8 hosts each. The maximal degradation is about 25\%, even for this small cluster of 32 nodes. The full bars on the single tenant runs demonstrate we normalize each run condition separately.}
\label{fig:job_performance_on_cluster} \vspace{-0.3cm}
\end{figure}

\RM{
We also run stencil application on the 32 nodes cluster. This MPI application runs cycles of computation and communication on virtual x, y or z axis. We measure the time to complete 100 compute/communicate iterations by the first job.
The jobs start one after the other with some delay, such that the resulting measurement show a gradual increase of the first job iteration time due to the growing number of jobs interfering.
The results are plotted in ~\fig{fig:meas_iter_time} which shows a degradation of $43\% = 0.215/0.15$ in the presence of 4 parallel jobs.
Note that on larger systems where the job sizes are larger and many more jobs exist the expected impact on job run-time is larger.
\begin{figure}[]
\centering
\includegraphics[width= 0.9 \columnwidth]{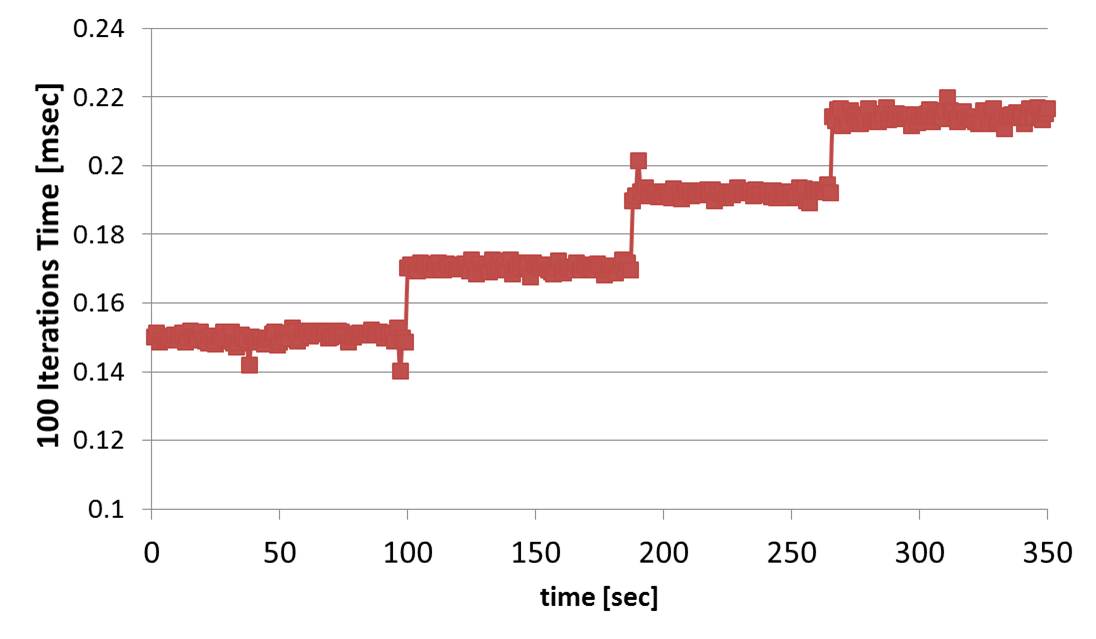}
\caption{MPI stencil computation app run-time, on a 32 nodes 10GB/s InfiniBand cluster, degrades by 43\% with the gradual start of 3 other similar apps.}
\label{fig:meas_iter_time}
\end{figure}
}

\T{Tenant interference in scaled-up simulations.} We now evaluate the impact of cloud size. As the number of tenants and their sizes grow, we would expect an increased inter-tenant friction, and therefore a degraded application performance in the presence of concurrent tenants.
We simulate the effect of the concurrent tenant traffic on a cloud of 1,728 hosts for 8 and 32 randomly-placed tenants, each of 216 and 54 hosts respectively.
We measure the average relative performance of a tenant, defined as the ratio of its performance when running concurrently with all other tenants by its performance when running alone.
We show the impact of inter-tenant friction on scientific-computing applications as well as on MapReduce.
For the scientific-computing benchmark, we select stencil codes, which are parallel programs that break the problem space (mainly 3-dimensional) into sub-spaces, apply the same procedure to each sub-space and exchange data mostly with neighboring sub-spaces.
\TR{This scheme is common to many scientific programs, and especially those solving partial differential equations, such as weather prediction and flow dynamics.}
The computation time is again kept constant while the communication time changes with the increasing message size. For a single tenant with 32KB messages, the communication time represents roughly 4/9 of the total time.

\fig{fig:simulated_runtime} shows how the relative performance of each tenant decreases as the number of tenants and the message size increase. For instance, for 32 concurrent tenants exchanging 32KB messages, the performance degrades by 45\% compared to a tenant running alone (equivalently, providing isolation from concurrent tenants would more than double the performance).
This significant loss of performance happens despite a modest message size of 32KB, and presents a large source of potential run-time variability.
Note that the degradation of performance is clearly a result of network contention, since each job runs on dedicated hosts.
MapReduce (simulated at similar conditions) experiences a smaller impact than stencil applications.
Interestingly, the smaller interference from other tenants is a result of higher self-contention: due to the Shuffle all-to-all traffic pattern, there is network contention even when MapReduce runs alone. Stencil applications suffer less from self-contention because their traffic matrix is less dense.
\begin{figure}[t]
\centering
\includegraphics[width= 0.9 \columnwidth]{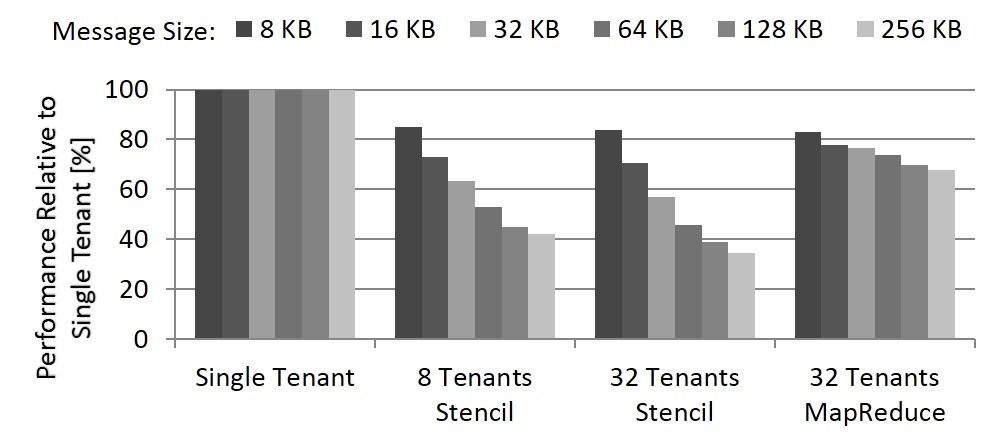}
\caption{Simulated relative performance for 8 and 32 tenants on a cloud of 1,728 hosts, with Stencil scientific-computing applications or MapReduce-based applications.
\TR{The relative performance to a single tenant degrades as the traffic volume and the number of concurrent tenants increase.}}
\label{fig:simulated_runtime}
\end{figure}
\RM{
Our second set of simulations illustrates tenant interference on a partition-aggregate traffic pattern, which is characteristic of distributed database queries run by many Web2.0 services like Facebook~\cite{zats_detail:_2012,alizadeh_data_2010,vamanan_deadline-aware_2012}.
We simulate such a traffic pattern on the same cluster, assuming each of the 32 tenants splits its hosts equally between servers and clients. The query arrivals follow a Poisson process with a controllable rate. Each query is sent to all servers in parallel.

\fig{fig:sim_pa_sat} shows the percentage of late queries not meeting a 10-msec deadline.
The steep increase of late queries happens at about 10,450 queries per second for the 32 concurrent tenants, versus 13,600 queries per second for a single tenant. The network link sharing resulted in a degradation of about 30\% in the effective query rate.
\begin{figure}[t]
\centering
\includegraphics[width= 1.0 \columnwidth]{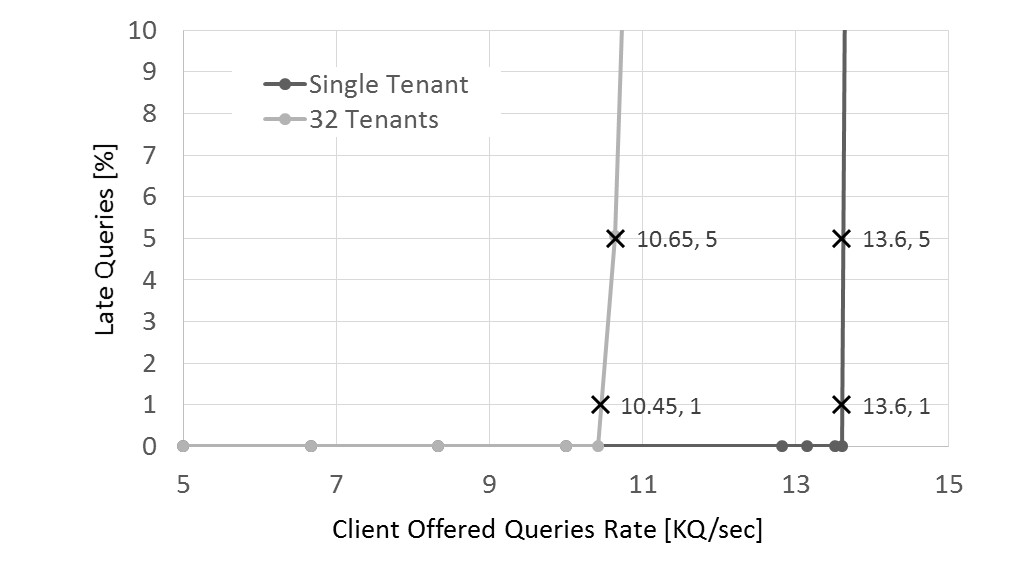}
\caption{Simulated distributed database tenants placed randomly on 1,728 nodes cluster.
The percentage of queries not meeting a 10msec deadline vs. offered query-rate show steep saturation.
}
\label{fig:sim_pa_sat}
\end{figure}
} 
\ORM{ 
\begin{figure}[t]
\centering
\includegraphics[width= 0.9 \columnwidth]{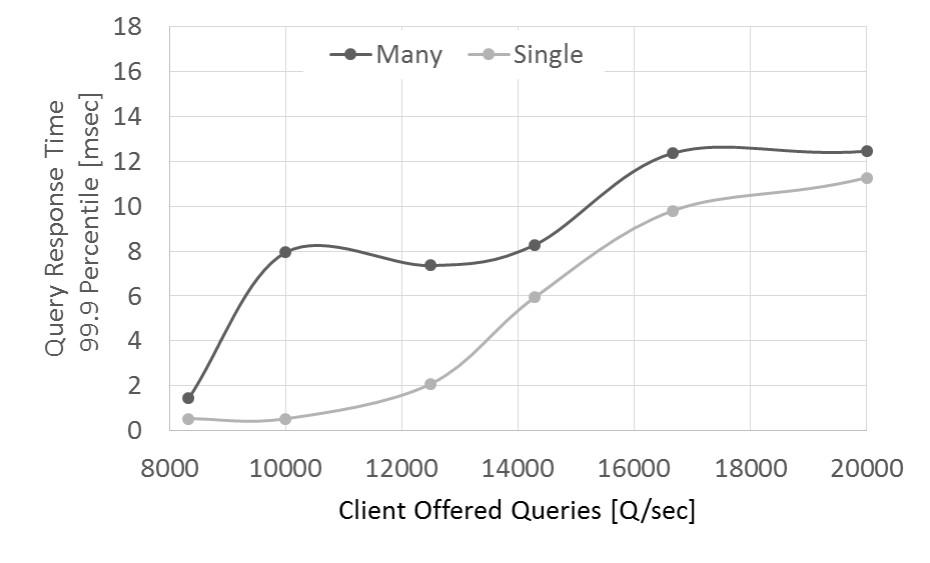}
\caption{Simulated distributed database queries 99.9 percentile of query latency for single and 32 tenants. In the presence of other tenants the query latency at 10Kquery/sec is 10 times larger if than running alone.}
\label{fig:sim_pa_lat}
\end{figure}
} 

We further want to confirm that similar results are obtained for a lossy Ethernet network.
We simulate a 32-node Ethernet cluster employing ECMP routing and DCTCP~\cite{alizadeh_data_2010},
using an INET~\cite{varga_overview_2008} simulator enhanced with a specially-implemented DCTCP plugin. We simulate 32 nodes and not 1,728 nodes because this simulator is less scalable.
There are only two tenants: The first is a regular 8-node tenant implementing MapReduce, of random Map and Reduce times  and variable Shuffle data size (producing a similar ratio of communication time to total time). The second is an 8-node adversarial aggressor tenant. Each adversarial node continuously generates 1MB messages, sent in parallel to all its other nodes.
We intentionally keep half the nodes unused to illustrate the detrimental impact of other tenants even in an over-provisioned cluster.
~\fig{fig:inet_shuffle} presents the relative performance of MapReduce in the presence of the adversarial tenant as compared to its performance when running alone. The worst relative performance is obtained for messages of 128KB, with a degradation of 25\% even in such a small and over-provisioned cluster. We suspect that the increase in the last value with 256KB message results from an artifact of DCTCP.
\begin{figure}[t]
\centering
\includegraphics[width= 0.9 \columnwidth]{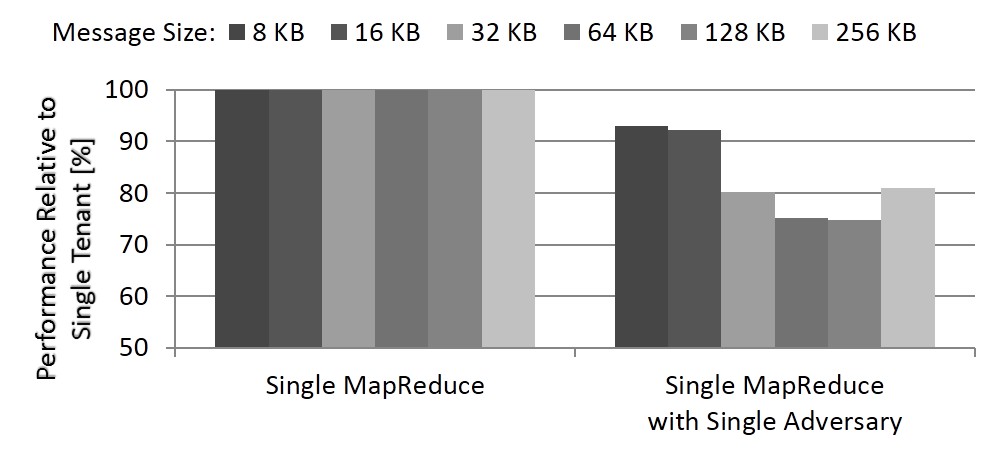}
\caption{Simulated relative performance of an 8-node MapReduce tenant on a 32-node Ethernet cluster running DCTCP. An adversarial 8-node tenant degrades the performance by 25\%.}
\label{fig:inet_shuffle}
\end{figure}

\section{LaaS Architecture}\label{sec:architecture}
A typical cloud architecture depicted in~\fig{fig:architecture} consists of (a) a \emph{front-end interface} for tenants to register their requests, (b) a \emph{ scheduler} that decides when and how to service these requests and can allocate hosts to tenants (e.g., an OpenStack Nova scheduler and a Heat application setup), and (c) a \emph{network controller} that performs the network setup (e.g., an OpenStack Neutron and an SDN back-end).
In this section, we introduce a \emph{LaaS cloud architecture} that enhances this architecture by enabling the allocation of  tenant-exclusive hosts and links.

Specifically, we propose to extend the \emph{scheduler} with link allocation functionality (on top of the host allocation), and enhance the \emph{network controller} by adding network routing rules to enforce the link allocation.
\fig{fig:architecture} emphasizes these two extensions by bold lines on 
an abstract cloud management software architecture.

\T{Scheduler.} We require the scheduler to provide each new tenant with an exclusive set of \emph{dedicated hosts} and \emph{dedicated links}.
As in bare-metal allocation, a tenant may request a given number of \emph{dedicated hosts}, which may be further refined by requirements of memory, accelerators or number of cores. In our implementation, we assume homogeneous hosts.
In addition, the scheduler provides each new tenant with a set of \emph{dedicated links} that form a tenant sub-topology, which will guarantee full bandwidth for any admissible traffic matrix of the tenant, i.e. will provide the tenant with the same bandwidth as in its own private data center.

In the LaaS architecture, we assume that the scheduler employs an online algorithm, by successively processing one new tenant request at a time. Each new tenant may be either accepted to the cloud, or denied due to the unavailability of a sub-network that can provide enough dedicated hosts and links. In any case, the scheduler does not migrate already-running tenants.
This could be relaxed if we want to allow global optimization of host placements, by running tenants over virtual machines (VMs) and allowing migrations~\cite{jiang_joint_2012,wen_virtualknotter:_2012,VMWARE_virtual_2011}. But then, tenant run-times would be impacted by the arrival of new tenants, which is precisely what we want to avoid.

\begin{figure}[]
\centering
\includegraphics[width= \columnwidth]{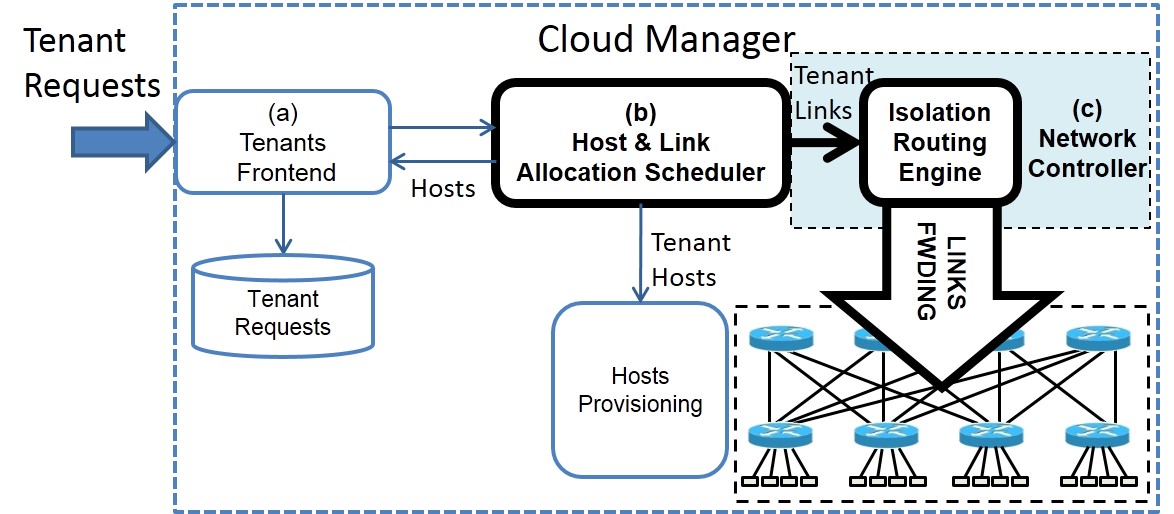}
\caption{Cloud management system architecture, with LaaS extensions in bold.}
\label{fig:architecture} \vspace{-0.3cm}
\end{figure}

\ORM{ 
\I{By moving links, can we really pack more links? If not, can we delete the following paragraph?}
The link allocation problem is different and can be solved offline as a global optimization problem.
The reason is that routing modifications to previously accepted tenants can be conducted  in a gradual manner that preserves the connectivity.
But out of order delivery of packets, during the routing changes, is much harder to guarantee~\cite{reitblatt_consistent_2011}.
} 

\T{Network controller.} As depicted in \fig{fig:architecture}, we require the information of the allocated links to be provided by the scheduler to the network devices.
This information should be used to adjust the network forwarding and routing to provide tenant isolation.
This task fits SDN networks, but may also be implemented in other network architectures like TRILL~\cite{perlman2008transparent}.
There are several different ways to implement such an isolation-aware network controller.
At one extreme, which requires switch-virtualization hardware support, a master controller may configure the underlying switches to be split into multiple virtual switches~\cite{sherwood_flowvisor:_2009}.
Then each tenant may incorporate its own SDN controller, which can then only discover its own isolated sub-topology.
Another approach is to let a single SDN controller do all the work and enhance all the routing engines to work on sub-topologies.
We rely in our implementation on an off-the-shelf InfiniBand SDN controller with a capability of defining sub-topologies and routing packets in an isolated manner (L2 forwarding).
This feature, known as Routing Chains, is described in~\cite{mellanox_ofed_um_2014}. This isolated-routing feature could also be implemented by Ethernet SDN controllers like OpenDaylight.

\section{LaaS Algorithm}\label{sec:analysis}
In this section, we describe online algorithms for \emph{tenant placement} and \emph{link allocation} in the LaaS scheduler.
Online placement algorithms require the existing tenant placement to be maintained when a new job is placed, and therefore do not move existing tenants. Similarly we provide online link-allocation algorithms to avoid any traffic interruption when a new tenant is introduced. The algorithm we describe provably guarantees that a tenant will obtain a dedicated set of hosts and links, with the same bandwidth as in its own private data center.
The algorithm relies on the required properties of the placement to trim the solution space and achieve fast results.

We first study 2-level fat-trees, and then generalize the results to 3 levels.
We first present a \emph{Simple} heuristic algorithm, and then extend it with a \emph{LaaS} algorithm that achieves a better cloud utilization.

\subsection{Isolation for 2-level Fat Trees}\label{sec:two_level_isolation}
Consider a 2-level full-bisectional-bandwidth fat-tree topology, i.e. a Full Bipartite Graph between leaf switches and spine switches, as in~\fig{fig:before_n_after} above.
For brevity we denote Full Bipartite Graphs that make the fat-tree connections between switches at levels $lvl_i$ and $lvl_{i+1}$: $FBG_i$. 
It is composed of $r$ leaf switches, denoted $L_i$ for each $i \in [1,r]$, and $m$ spine switches.
Each leaf switch is connected to $n \leq m$ hosts as required to meet the rearrangeably non-blocking condition for fat-trees~\cite{jajszczyk_nonblocking_2003}.

\ST{Problem definition}
Given a pre-allocation of tenants (with pre-assigned links and hosts), when a new tenant arrives with a request for $N$ hosts, we need to find: \\
\emph{(i) Host placement:} Find which free hosts to allocate to the new tenant, 
    i.e. allocate $N_i$ free hosts in each leaf $i$ such that $N = \sum\limits_{i=1}^r{N_i}$. \\
\emph{(ii) Link allocation:} Find how to support the tenant traffic, i.e. allocate a set $S_i$ of spines for each leaf $i$, such that the hosts of the new tenant in leaf $i$ can exclusively use the links to $S_i$, and the resulting allocation can fully service any admissible traffic matrix.

We want to fit as many arriving tenants as possible into the cloud such that their host placement and link allocation obey the above requirements, and without changing pre-existing tenant allocations.

\ST{Simple heuristic algorithm} 
We first introduce a \emph{Simple} heuristic algorithm, as basis for the discussion of our algorithm.
It relies on a property of fat-trees and minimum-hop routing: if a single tenant is placed within a sub-tree, then traffic from other tenants will not be routed through that sub-tree. Note that for 2-level fat-trees a sub-tree is a leaf switch.

Let $N$ denote the number of tenant hosts, and $n$ the number of hosts per leaf. The \emph{Simple} heuristic simply computes the minimal number $s$ of leaf switches required for the tenant: $s = \ceil{{{N}/{n}}}$. Then, it finds $s$ empty leaf switches to place the tenant hosts in. Finally, if $s>1$, it allocates all the up-links leaving the $s$ leaf switches; else, no such links are needed.

\fig{fig:simple} illustrates the \emph{Simple} algorithm, showing how tenant $T_1$ obtains a placement for $N=6$ hosts. 
First, $s=\ceil{{6}/{4}}=2$. Assuming $T_1$ arrives first, the two left leaves are available when it arrives, and they are used to host $T_1$. Also, all the up-links of these 2 leaf switches are allocated to $T_1$. When it arrives, tenant $T_2$ is similarly allocated the two right leaves and their up-links.

In the general case, any placement obtained by \emph{Simple} supports any admissible traffic pattern. This is because the dedicated sub-network of the tenant is a single leaf switch if $s=1$, and a 2-level fat-tree if $s>1$, which is a folded-Clos network with $m \geq n$.
It is well known that such a topology supports any admissible traffic pattern, because it meets the rearrangeable non-blocking criteria and the Birkhoff-von Neumann doubly-stochastic matrix-decomposition theorem~\cite{jajszczyk_nonblocking_2003}.

\begin{figure}[]
\centering
\includegraphics[width=0.7 \columnwidth]{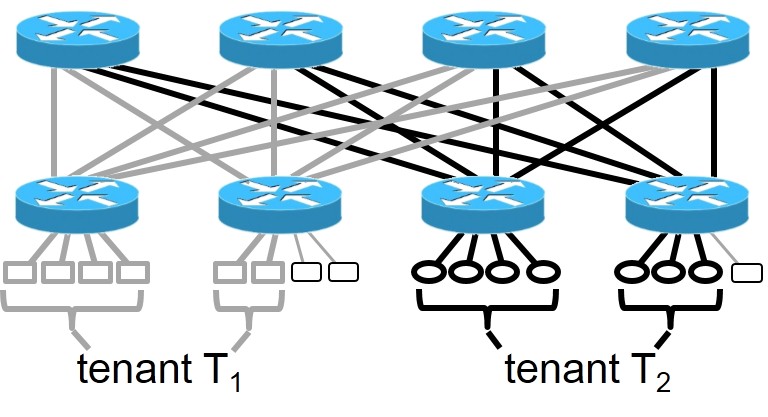}
\caption{Two tenants of sizes 6 and 7 hosts placed by the \emph{Simple} heuristic, where each tenant fills a number of complete sub-trees.}
\label{fig:simple} \vspace{-0.3cm}
\end{figure}

\ST{Extended Simple Heuristic}
It is possible to allow a single tenant with hosts within a sub-tree to span across multiple sub-trees. The same argument used for the simple case holds for the extended case since only the traffic of the single tenant, leaving the sub-tree, is crossing the top level of the sub-tree. Thus isolation is maintained. Since the entire set of links at the top layer must match the number of hosts within that sub-tree the obtained topology supports any admissible traffic matrix.

For example~\fig{fig:simple_ext} shows how tenant $T_3$ occupies a part of a leaf sub-tree which is shared by tenant $T_1$ extending out of that sub-tree. No traffic other that of $T_1$ would need to leave the same sub-tree and thus all the top links in that tree are allocated to tenant $T_1$.
\begin{figure}[]
\centering
\includegraphics[width=0.75 \columnwidth]{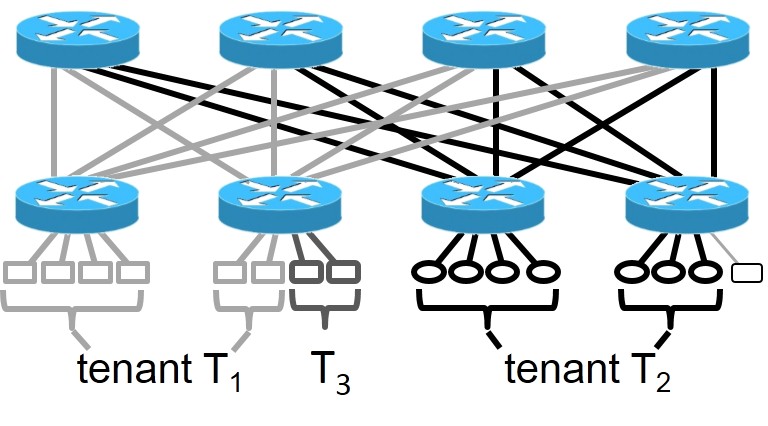}
\caption{Three tenants placed by the \emph{Extended Simple} heuristic.
Note that $T_1$ and $T_3$ are sharing the same sub-tree (leaf) but only one of them, $T_1$ is allowed to expand out of the shared sub-tree.}
\label{fig:simple_ext}
\end{figure}


\ST{LaaS placement analysis} 
This section describes a required condition on placement and sufficient condition on link allocation that are key to make the LaaS algorithm correct and efficient.
The placement condition requires the allocation of $N$ tenant hosts as $Q$ leaves of $D$ hosts and optionally additional leaf of $R \mid R < D$ hosts such that $N=QD+R$.
The sufficient link allocation condition requires the links of $R$ spines connecting to the $Q$ leaves and the optional single leaf of $R$ hosts. A subset of size $D-R$ of these spines should connect just to the $Q$ leaves.

Consider a single leaf $i$ with $N_i$ tenant hosts.
In the analysis below, we make the following simplifying assumption: on every leaf switch, the number of leaf-to-spine links (and the corresponding number of spines) allocated to a tenant equals the number of its allocated hosts:
\be \label{eq:NiSi} \abs{S_i} = N_i.\ee
Our simplifying assumption is based on the following intuition. On the one hand, for tenants occupying several leaves, if $\abs{S_i} < N_i,$ we may not be able to service all admissible traffic demands (since we may have up to $N_i$ flows that need to exit leaf $i$, but only $\abs{S_i}$ links to service them).
On the other hand, allocating $\abs{S_i} > N_i,$ is wasteful, because the number of remaining spine switches would then be less than the number of available hosts, and therefore future tenants spanning more than one leaf may not be able to obtain enough links to connect their hosts.

Without loss of generality, we also make a notational assumption that the $N_i$'s are sorted such that $0 < N_1 \leq N_2 \leq \dots \leq N_t$, where $t$ is the number of leaves connected to hosts allocated to the tenant.

We will now see that our assumptions lead (by a sequence of lemmas) to a simple rule that greatly simplifies the possible placements that need to be evaluated by our LaaS scheduling algorithm.

\begin{lemma}\label{lem:2}
The number of common spines that connect two leaves must at least equal their minimal number of allocated hosts:
\begin{equation}
\forall i < j \in [1,t]: N_i = \min(N_i, N_j) \leq \abs{S_i \cap S_j}
\end{equation}

\end{lemma}
\PR{
\bp
Consider a traffic permutation among the tenant hosts. There are up to $N_i$ full-link-capacity host-to-host flows going from $L_i$ to $L_j$ (or back).
Since each flow has to use a different link and each link goes to a different spine switch, we will need at least $N_i$ common spine switches in $\abs{S_i \cap S_j}$.
\ep
} 

\begin{lemma}\label{lem:3}
The number of common spines that connect two leaves to a third must at least equal the minimal number of allocated hosts, either in the union of the first two leaves or in the third, i.e.
$\forall i,j,k \in [1,t]: \min(N_i+N_j, N_k) \leq \abs{S_i \cup S_j}$.
\end{lemma}
\PR{
\bp
Let $c=\min(N_i+N_j, N_k)$.
There are at most $c$ flows going from $L_k$ to either $L_i$ or $L_j$ (or back).
Since each flow has to use a different link and each link goes to a different spine switch, we will need at least $c$ spines in the union ${S_i \cup S_j}$ of the spines connected to the two leaves.
\ep
} 

\begin{lemma}\label{lem:5}
The number of allocated hosts in any leaf cannot exceed the number in the union of any two other leaves, i.e.
$\forall i\neq j \neq k \in [1,t]: N_i, N_j, N_k > 0 \rightarrow N_i+N_j \geq N_k$
\end{lemma}
\PR{
\bp
Assume the contrary: $N_i+N_j < N_k$.
There are only two cases: $N_i \leq N_j < N_k$ or $N_j \leq N_i < N_k$.  W.l.o.g., we assume the first.
If so, $\min(N_i+N_j, N_k) = N_i+N_j$.
By \lem{lem:2}, to enable connectivity between $N_i$ and $N_j$, they must have at least $N_i$ spines in common: $\abs{S_i \cap S_j} \geq N_i$.
Substituting the above into \lem{lem:3} we obtain:
$\forall i,j,k \in [1,t]: \min(N_i+N_j, N_k) =  N_i+N_j \leq \abs{S_i \cup S_j}=\abs{S_i}+\abs{S_j}-\abs{S_i \cap S_j}$.
But since $N_i=\abs{S_i}$ and $N_j=\abs{S_j}$ in LaaS by \eq{eq:NiSi}, we get
$0 \leq -\abs{S_i \cap S_j}$. But $S_i \cap S_j$ is non-empty because otherwise traffic from hosts in leaf $i$ to hosts in $j$ wouldn't be able to pass.
So we get a contradiction, thus $N_i+N_j \geq N_k$.
\ep
} 

\ST{Necessary host placement} We will now provide two theorems showing necessary and sufficient conditions to get the LaaS conditions of tenant traffic isolation and support for any admissible traffic matrix. Interestingly, the first theorem requires \emph{necessary conditions on the host placement}, while the second theorem provides  \emph{sufficient conditions on the link allocation}. We continue to assume throughout the rest of the paper that $\abs{S_i} = N_i$ for all $i$, and $N_1 \leq N_2 \leq \dots \leq N_t$.

\begin{theorem}\label{thm:RNBISOL_PLACEMENT}
A necessary condition for LaaS  is \be N_1 \leq N_2 = N_3 = \dots = N_t, \ee implying that all leaf switches of a tenant should hold the exact same number of hosts except for a potential smaller one.
\end{theorem}
\PR{
\bp
We show that $N_2=N_t$.
By \lem{lem:2}, $L_1$ and $L_2$ must have at least $N_1=\abs{S_1}$ spines in common, i.e. $S_1 \subseteq \para{S_1 \cap S_2}$. Therefore, $S_1$ is a subset of $S_2$, 
so $\abs{S_1 \cup S_2}=\abs{S_2} = N_2$.
By \lem{lem:5}, when $i=1$, $j=2$ and $k=t$, $N_1+N_2 \geq N_t$ thus $\min(N_1+N_2,N_t) = N_t$.
So, when $N_t$ flows are sent from $L_t$ to $L_1$ and $L_2$, we must have at least $N_t$ common spines: $\abs{S_1 \cup S_2} = N_2 \geq N_t$.
But since $N_2 \leq N_t$, it follows that $N_2=N_t$.
\ep
} 
Given \thm{thm:RNBISOL_PLACEMENT}, the tenant placement should follow the form: $N=Q\cdot D+R$, where $Q$ is the number of repeated leaves with $D$ hosts each, and we optionally add one unique leaf with a smaller number of hosts $R$. This notation follows the \underline{D}ivisor, \underline{Q}uotient and \underline{R}emainder of $N$.
This result is useful because it greatly simplifies the solution of the host placement problem defined above.

\fig{fig:placement_rule} demonstrates this result. It shows $Q$ leaf switches of $D$ hosts each, and optionally another leaf switch of $R<D$ hosts. We denote by $S^{D}$ the set of spines connected by allocated links to the $Q$ leaves of $D$ hosts, and by $S^{R}$ those that connect via allocated links to the optional leaf of $R$ hosts.

\begin{figure}[]
\centering
\includegraphics[width= 0.9 \columnwidth]{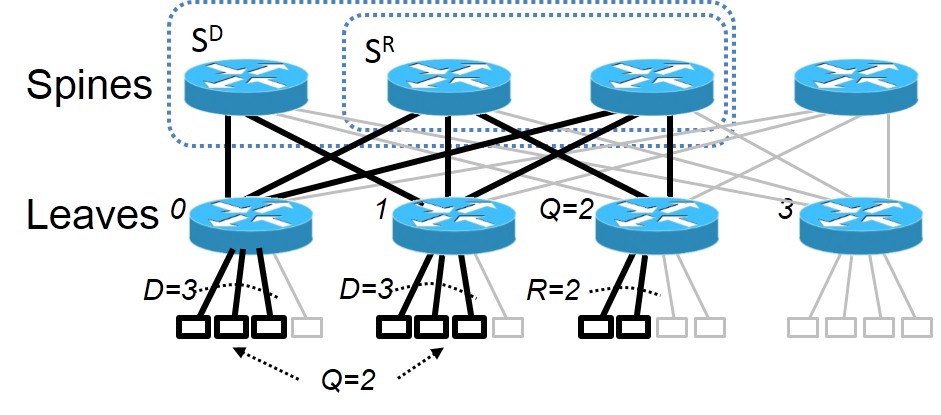}
\caption{A tenant of $N=8=Q \cdot D + R$ hosts. To implement LaaS, there must be $Q$ leaves of $D$ hosts and optionally one leaf of $R < D$ hosts.}
\label{fig:placement_rule} \vspace{-0.3cm}
\end{figure}

\ST{Sufficient link allocation} We can now prove sufficient conditions on the link allocation to satisfy LaaS.
\begin{theorem}\label{thm:RNBISOL_LINKS}
A sufficient condition for LaaS is that the link allocation satisfies $\forall i \in [1,Q]: S_i = S^{D}$ and if $R>0 : S^{R} \subset S^{D}$, i.e. all the allocated leaf up-links of a given tenant go to the exact same set of spine switches (or a subset of it for the remainder leaf).
\end{theorem}
\PR{
\bp
For the case $R=0$, the link allocation above means there is a group of $D$ spine switches that connect to all leaf switches. Thus the tenant sub-topology reduces to an \emph{Full Bipartite Graph} $(FBG)$ with $m'=D$ spine switches and $n'=D$ hosts per leaf.
Since $m' = n'$ such topology is rearrangeable non-blocking folded-Clos which is known to support any admissible traffic matrix as mentioned above.

For the case of one additional leaf $L_{j_R}$ of $R$ hosts, we provide a constructive method for routing arbitrary permutations. We consider the $FBG$ sub-topology formed by the tenant hosts and links,
where $L_{j_R}$ connects to all $S^{D}$ spines. 
For this topology $m'=n'=D$ and $r'=Q+1$.
Again, $m'=n'$ so it is guaranteed by the rearrangeable non-blocking theorem that every full permutation of $n' \cdot r'$ flows is route-able.
Routing is symmetric with respect to the spine switches.
Moreover, to avoid congestion, each spine needs to carry exactly 1 flow from each leaf and 1 flow to each leaf.
So any full permutation of our original topology where $L_{j_R}$ has only $R$ flows will be $D-R$ flows short.
We extend these flows with $D-R$ flows going from $L_{j_R}$ to $L_{j_R}$.
Since these flows share the same leaf switch they must be routed through $D-R$ different spines.
After completing the full permutation routing, and since all spines connect to all leaves, we swap between each spine that carries one of the added $D-R$ flows with a spine that is not included in $S^{R}$.
As the links allocated to the extra flows are not needed, any permutation is fully routed by the original topology.
\ep
} 
\ST{A necessary host allocation is not sufficient}
The above theorems provide us with guidelines for implementing LaaS. 
We now show that due to previous tenant allocations, a host placement as in
\thm{thm:RNBISOL_PLACEMENT} is not always sufficient to provide a needed link allocation as in~\thm{thm:RNBISOL_LINKS}. This is why \thm{thm:RNBISOL_LINKS} proves essential. If the link allocation cannot be found for a specific placement our algorithm will need to search for another host allocation.
\begin{lemma}\label{lem:pl_not_suff}
A host placement that meets~\thm{thm:RNBISOL_PLACEMENT} does not guarantee the existence of a link allocation that meets~\thm{thm:RNBISOL_LINKS}, and therefore does not guarantee LaaS.
\end{lemma}
\PR{
\bp
\begin{figure}[]
\subfigure[Placement] {
\includegraphics[width=0.5 \columnwidth]{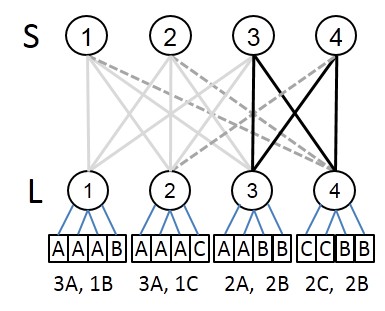}
\label{fig:la_feas_a}
}
\subfigure[Link Allocation] {
\includegraphics[width=0.4 \columnwidth]{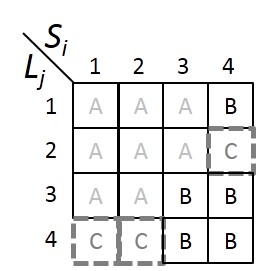}
\label{fig:la_feas_b}
}
\caption{Illustration that a simple host placement is not sufficient, and a joint host placement and link allocation is necessary for LaaS. (a) All tenants satisfy the host placement necessary conditions
, e.g. the placement of C is $3=Q\cdot D + R=2\cdot1+1$. A and B support any admissible traffic matrix by the sufficient link allocation conditions
.
(b) However, the link allocation for  C is impossible.  There is no way to find a common set of spines with free ports.}
\label{fig:disprove_link_alloc_feas} \vspace{-0.3cm}
\end{figure}
We prove~\lem{lem:pl_not_suff} by the example provided in \fig{fig:disprove_link_alloc_feas}.
Three tenants are shown placed according to the provided heuristic of the previous section: A has $8=2\cdot 3+2$ hosts, B has $5=2\cdot 2+1$, and C has $3=1 \cdot 2+1$.
We track allocated up-links of the leaf switches in a matrix where rows represent the leaf switches and columns represent the spines each port connects to.
As can be observed, there is no possible link allocation for tenant C, since the leaves it is placed on do not have free links connected to any common spine.
There is no link allocation possible for C even though it was placed according to the conditions of ~\thm{thm:RNBISOL_PLACEMENT}.
The online link allocation algorithm for C (after A and B are placed) cannot allocate the links.
In fact, even an offline version of link allocation - reassigning the links of A and B - cannot solve the problem once the placement of A and B does not change.
\ep
} 
According to~\lem{lem:pl_not_suff}, some tenant requests may be denied because the scheduler cannot find a proper link allocation. Thus any LaaS algorithm has to validate the feasibility of a link allocation for each legal host placement.

\subsection{Isolation for 3-level Fat Trees}
So far we have discussed the LaaS allocation for 2-level fat-trees. We now extend the results to 3-level fat-trees, which form the most common cloud topology~\cite{al-fares_scalable_2008,andreyev_introducing_2014}.
We use the notation of Extended Generalized Fat Trees (XGFT)~\cite{ohring_generalized_1995}, which defines fat-trees of $h$ levels and the number of sub-trees at each level: $m_1, m_2,\dots,m_h$. \TR{and the number of parent switches at each level: $w_1,w_2,\dots,w_h$.}

We consider three approaches to this problem: a \emph{Simple} heuristics, a \emph{Hierarchical} decomposition, and an \emph{Approximated} scheme. We conclude with a description of the final \emph{LaaS} algorithm that we implemented, relying on the \emph{Approximated} scheme.

\ST{Simple heuristic for 3-level fat-trees}
The \emph{Simple} algorithm described in sub-section
'Simple heuristic algorithm'
is easily extended to any fat-tree size.
For an arbitrary XGFT, first define  the number of hosts $R_l$ under a sub-tree of level $l$: $R_0 = 0$, and $R_l=\prod_{i=1}^{l}{m_i}$.
Given a tenant request for $N$ hosts, \emph{Simple} first determines the minimum level $l_{min}$ of the tree that can contain all $N$ tenant hosts:
\begin{equation}
l_{min} =\min \brac{l |
    \para{R_{l-1} < N}
\wedge
    \para{R_{l} \geq N}
}
\end{equation}
and the number $s$ of required sub-trees of level $l_{min}$: $s = \ceil{{N}/{R_{l_{min}-1}}}$.
Then, it places the tenant hosts in $s$ free sub-trees of level $l_{min}$. It also allocates to the tenant all the links internal to these $s$ sub-trees; and if $s > 1$, it allocates as well all the links connecting the sub-trees to the upper level.

It is clear that the \emph{Simple} heuristic algorithm, by rounding up the number of nodes, trades off cluster utilization for simplicity, non-fragmentation, and greater locality with lower hop distances.
As we show in the evaluation section, the utilization obtained by this algorithm is low, making it potentially unacceptable to cloud vendors, so we keep looking for a better one.

{\ST{Hierarchical decomposition} 
In this section we describe how LaaS can be provided to a 3-level fat-tree using a hierarchical decomposition approach following the recursive description of fat-trees in~\cite{petrini_k-ary_1997}.

\fig{fig:decomp} shows an example of 3-level fat-tree.
We denote the switches on the tree by their levels (from bottom up) $lvl_1$, $lvl_2$ and $lvl_3$.
We show that for a LaaS link allocation to be feasible, the condition of~\thm{thm:RNBISOL_PLACEMENT} needs to hold not only for each 2-level sub-tree but also for each $lvl_2$ - $lvl_3$ Full Bipartite Graph ($FBG_2$) at the top of the tree.
One of these $FBG$s is highlighted in~\fig{fig:decomp}.

As we showed in the previous sections, since the tenant traffic pattern may be completely contained within each 2-level tree, host allocation in each 2-level tree must adhere to~\thm{thm:RNBISOL_PLACEMENT}.
So the number of tenant hosts within the 2-level sub-tree $j$ must be of the form $N_{j}=Q_j\cdot D_j + R_j$.
Note that an allocation that fits in a single leaf switch also follows this scheme with $Q_j=1$.

\fig{fig:decomp} depicts a \thm{thm:RNBISOL_PLACEMENT}-compliant host allocation within each of the 2-level sub-trees. It follows the form: $N_{j}=Q_j\cdot D_j + R_j | j \in \brac{1 ... m_3}$.
Note that the link assignment within the 2-level sub-trees must also adhere to \thm{thm:RNBISOL_LINKS} such that $S^R_j \subset S^D_j$.
Consequently, the maximum number $U_j$ of flows leaving the 2-level sub-tree from switch $s$ can be either $0$ in case $s \notin S^D_j$, $Q_j$ in case $s \in S^D_j \backslash S^R_j$, or $Q_j+1$ if $s \in S^R_j$.

When we consider the conditions required for the highlighted $FBG_2$ to support any admissible traffic pattern, it is strikingly similar to the analysis we provided for the 2-level fat-tree.
For the 2-level tree we already proved that in order to support any admissible traffic pattern,
the sequence of $U_j$ values must meet the rule $U_1 \leq U_2 = U_3 = \dots = U_{m_3}$.
Applying the same to the 3-level tree we obtain a requirement for the assignments of $U_j$ on each of the $FBG_2$.
However, each one of the $FBG_1$ (there are $m_3$ such 2-level sub-trees) could select a different set of $S^D_j$ and $S^R_j$. This means that a solution could allow each 2-level sub-tree to select a different set of $FBG_2$ to carry its flows, as long as the above rule is maintained for each $FBG_2$.

Unfortunately the above rule still allows a vast amount of legal tenant-placement and link-allocation possibilities, which make the full 3-level fat-tree LaaS problem too hard to be solved in practical time even on high-end processors.
If we were to provide an optimal allocation we would conclude here that our problem is too hard.
But our task is not to find the \emph{optimal} solution, or even \emph{any} solution at a specific iteration.
Our target is to show that there is a simple enough algorithm that would be able to handle the online LaaS problem in reasonable time and with reasonable success rate such that the cluster utilization remains high and LaaS is guaranteed.
We do that by applying a restriction on the solution space of the hierarchical decomposition.

\begin{figure}[t!]
\centering
\includegraphics[width= 0.85 \columnwidth]{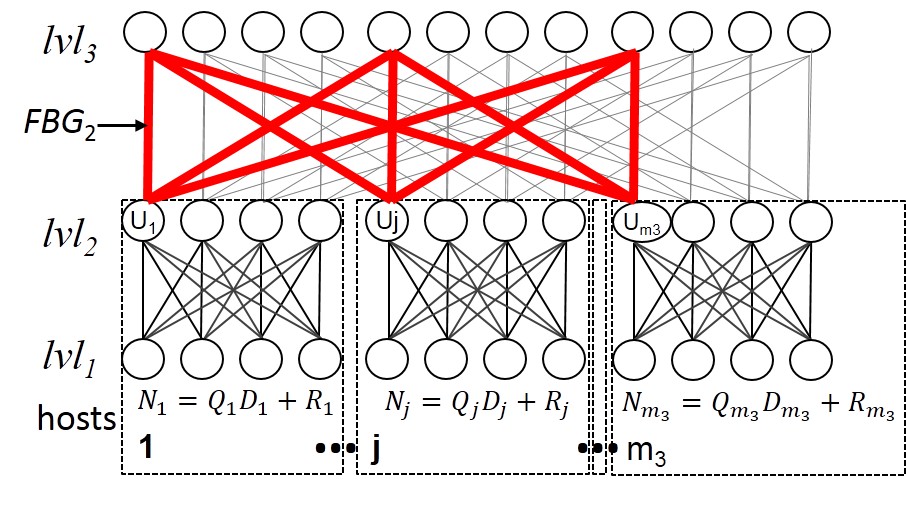}
\caption{A 3-level fat-tree showing the host allocation on each 2-level sub-tree matching \thm{thm:RNBISOL_PLACEMENT}.
One of the $lvl_2$ - $lvl_3$ Full Bipartite Graphs ($FBG_2$) is highlighted. We denote as $U_j$ the maximal number of flows injected into this $FBG_2$ from the $j^{th}$ 2-level sub-tree.}
\label{fig:decomp} \vspace{-0.5cm}
\end{figure}


\ST{Approximated algorithm} 
We provide a simpler algorithm that compromises cluster utilization in favor of reduction of the solution search space.
Our approximation requires the allocation to be symmetrical with respect to all the $FBG_2$, i.e. that the allocation on all the $FBG_2$ is identical and thus calculated just once.
So the solution must use the same number of flows $U_j$ leaving any one of the $lvl_2$ switches in the same 2-level sub-tree.
Note that any allocation where the number of tenant hosts $N_i$ connected to leaf switch $i$ does not include all the hosts on that leaf switch $N_i < m_1$, will not utilize all the links from that switch to the upper-level switches.
So only a subset of the $lvl_2$ switches in the same $FBG_1$ is going to pass traffic of that tenant.
Thus if we now consider the $lvl_2$ to $lvl_3$ traffic, not all $FBG_2$ will see the same $U_j$.
To avoid this we require that $D$ is either $0$ or $m_1$ for all 2-level sub-trees, except where the tenant fits within the same 2-level fat-tree and thus $U_j=0$.
As a consequence, if a tenant cannot fit within a single sub-tree, we round up its size to a multiple of $m_1$.
The host placement can now be performed in complete leaf switches of $m_1$ hosts.
For instance, if each leaf switch can hold $10$ hosts, and a tenant requests $N=267$ hosts, then we effectively allocate it $N' = m_1 \ceil{{N}/{m_1}}=270$ hosts.

Moreover, since the approximation in 3-level fat-tree allocates complete $lvl_1$ switches, it is equivalent to the 2-level LaaS problem: $lvl_1$ switches are equivalent to hosts, $lvl_2$ switches are like leaf switches and $lvl_3$ switches are like spines.
Thus the approximated 3-level fat-tree LaaS problem has to comply to the same conditions as for the 2-level tree.
We denote the allocation of full $lvl_1$ switches using a similar notation to the 2-level:
$Q'$ is the number of allocated 2-level sub-trees, each with $D'=Q$ leaves.
Optionally there may be one additional 2-level sub-tree with $R'$ allocated leaves.
$N' = \ceil{{N}/{m_1}} = Q'\cdot D' + R'$.

An example of such allocation for a tenant of 32 hosts on a 3-level fat-tree, with $m_1=4$ hosts per leaf, is provided in~\fig{fig:decomp_example}.
On the left $Q'=2$ sub-trees, the tenant uses $D'=3$ leaves and thus $U_1 = U_2 = 3$ for all $FBG_2$.
In addition a single unique sub-tree $r$ with $R'=2$ leaves is also allocated and thus $U_r = 2$ for all $FBG_2$.
So all the $FBG_2$ are thus identical.
Each one of them has to support $Q'$ $lvl_2$ switches of $D'=3$ flows and one $lvl_2$ switch with $R'=2$ flows.
These requirements meet the condition of~\thm{thm:RNBISOL_PLACEMENT} and thus may be feasible.

\begin{figure}[t!]
\centering
\includegraphics[width= 0.8 \columnwidth]{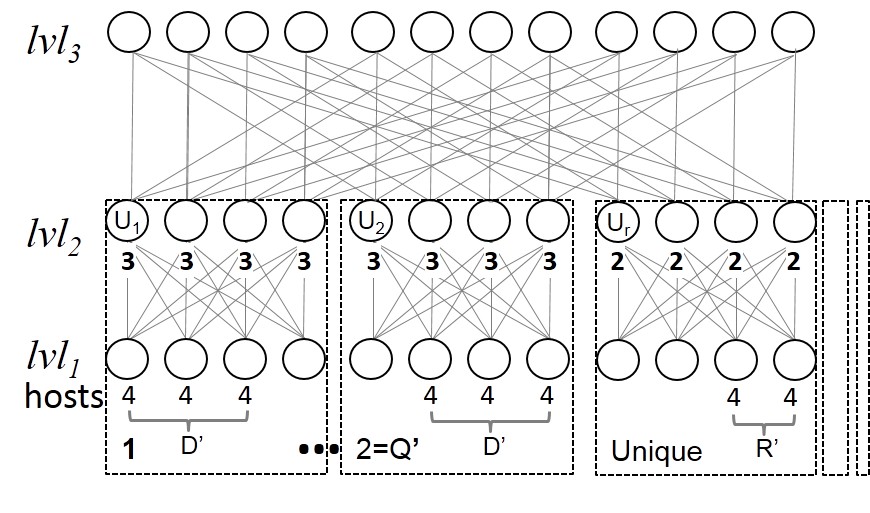}
\caption{An example of host placement with $N=32$ hosts on a 3-level fat-tree using the \emph{Approximated} method.
Using a notation similar to the 2-level fat-tree, this allocation is of the form: $Q'=2$, $D'=3$ and $R'=2$.}
\label{fig:decomp_example} \vspace{-0.25cm}
\end{figure}

\ST{LaaS algorithm} 
\begin{algorithm}[!t]
\scriptsize \caption{{\sc FLAP}($D,Q,R,l,l_e,r,\brac{ports},\brac{rl}$)}
\label{alg:alg1}
\begin{algorithmic}[1]
    \STATE {// find next $Q$ size leaf}
	\FOR{$i=l$ \TO $l_e$}
		\IF{$\abs{M\sbrac{i}} >= Q$}
		    \STATE $\brac{nPorts}$ = $\brac{ports} \cap M\sbrac{i}$
	   		\IF{$\abs{nPorts} \geq Q$}
	      		\STATE{$\brac{newRL} = \brac{rl} \cup i$}
	      		\IF{$r=D$}
	      			\STATE{// found all repeated leaves}
	          		\IF{findUniqueLeaf($R, l_s, l_e; \brac{nPorts} \brac{rl}$)}
	              		\STATE{$\brac{D_{PORTS}}=\brac{nPorts}$}
	              		\STATE{$\brac{D_L} = \brac{newRL}$}
	              		\STATE{return true}
	              	\ENDIF
	      		\ELSE
	      			\STATE {$j=i+1$; $s=r+1$}
	      			\IF{FLAP(D,Q,R,$j$,$l_e$,$s$,$\brac{nPorts}$,$\brac{newRL}$)}
	              		\STATE{return true}
	              	\ENDIF
	            \ENDIF
	       	\ENDIF
	    \ENDIF
	\ENDFOR
	\STATE return false
\end{algorithmic}
\end{algorithm}

\begin{algorithm}[!t]
\scriptsize \caption{{\sc LAAS}($N$)}
\label{alg:alg2}
\begin{algorithmic}[1]
	\STATE // Try 1 level allocation
	\IF{$N \leq m_1$}
		\FOR{$l=0$ TO $m_2\cdot m_3-1$}
			\IF {$FLAP(N,1,0,l,l,0,\brac{},\brac{}$)}
				\STATE return true
			\ENDIF
		\ENDFOR
	\ENDIF
	\STATE // Try 2 level allocation
	\IF{$N \leq m_1 \cdot m_2$}
		\FOR{$D=max(N,m_1)$ \TO $1$}
			\STATE $Q=\floor{\frac{N}{D}}$
			\STATE $R=N-Q\cdot D$
			\FOR{$l=0$ TO $m_3-1$}
				\IF {$FLAP(D,Q,R,l\cdot m_2,(l+1)\cdot m_2-1,0,\brac{},\brac{}$)}
					\STATE return true
				\ENDIF
			\ENDFOR
		\ENDFOR
	\ENDIF
	\STATE // Try 3 level allocation
    \STATE $U = \ceil{\frac{N}{m_1}}$
    \FOR{$D=max(U,m_2)$ \TO $1$}
		\STATE $Q=\floor{\frac{U}{D}}$
		\STATE $R=U-Q\cdot D$
		\IF{$Q \leq m_3$}
			\IF {$FLAP2(D,Q,R,0,m_3-1,0,\brac{}, \brac{}$)}
				\STATE return true
			\ENDIF
       	\ENDIF
	\ENDFOR
	\STATE return false
\end{algorithmic}
\end{algorithm}
We now want to implement our final \emph{LaaS} algorithm for concurrent host placement and link allocation in fat-trees. To do so, we rely on our \emph{Approximated} approach, and track the allocated up-links in a matrix similar to
~\fig{fig:link_alloc_tbl_a}.
The required set of leaves and links is of the form $N=Q\cdot D+R$.
As described in the sub-section
'LaaS placement analysis', 
in a general fat-tree, this translates to $R$ spines that connect to all the $Q+1$ allocated leaves and $D-R$ spines connected just to the $Q$ repeated leaves.
These requirements are equivalent to finding a set of $Q$ leaves that have $D$ free up-ports to a common set of spines, and a single leaf that has only $R$ free up-ports that form a subset of the spines used by the previous $Q$ leaves.

The search for $Q$ leaves with enough common spines is performed recursively. In the worst case, it may require examining all ${m_2 \choose Q}$ combinations. Our \emph{LaaS} algorithm returns the first successful allocation, so trying the most-used leaves first packs the allocations and achieves the best overall utilization results.

\begin{figure}[!t]
\centering
\subfigure[Link Allocation Table] {
\includegraphics[width=0.4 \columnwidth]{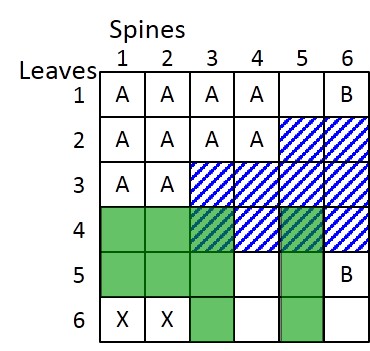}
\label{fig:link_alloc_tbl_a}}
\subfigure[Corresponding Topology] {
\includegraphics[width=0.51 \columnwidth]{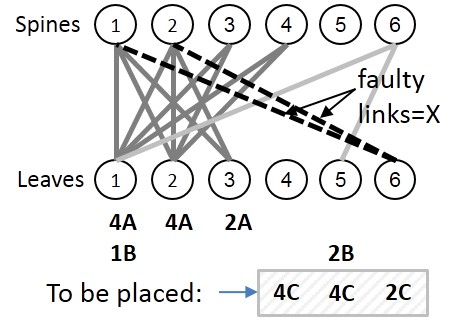}
\label{fig:link_alloc_tbl_b}}
\caption{Example of allocation with 2 potential placements. (a) Table of leaf up-links holding the link assignments of tenants A and B, as well as 2 faulty links X. (b) Corresponding topology.
The new tenant C of 10 hosts, arranged as $Q\cdot D+R =2 \cdot 4 + 2$, can be assigned one of two allocations.
In (a), the first link allocation is shown in solid, and the second with slanted lines.
}
\label{fig:link_aloc_tbl} \vspace{-0.3cm}
\end{figure}

\fig{fig:link_aloc_tbl} demonstrates the process of evaluating a specific $D, Q, R$ division. Consider a new tenant C of 10 hosts, arranged as 2 leaves of 4 hosts plus 1 leaf of 2 hosts. We show 2 possible placements: The first would use 4 hosts on leaves 4 and 5, and 2 hosts on another leaf 6. The second would use 4 hosts on leaves 3 and 4, and 2 hosts on another leaf 2. We also illustrate how we could take into account two faulty links in our link allocation if needed.

In the following section we describe the algorithm for mapping free leaves. 
The algorithm to perform the above example is provided in Algorithm~\ref{alg:alg1}.
The recursive function is assuming the availability of matrix $M\sbrac{l}$ of free ports on each leaf switch.
It is given the following constants: $D, R, Q$ and the start and end leaf switch indexes $l_s, l_e$.
The recursive function provides its current state on the recursion using the following variables:
$l$ represents the current leaf index to examine, $r$ the number of $Q$ size leaves that were already found, $\brac{ports}$ the set of ports that are possible for this allocation, $\brac{rl}$ the collected set of, so far, $Q$ size leaves.
Eventually the recursion provides the following results: $\brac{D_L}$ set of leaves with $Q$ hosts, $\brac{D_{PORTS}}$ the set of ports to be used by the $Q$ size leaves, $U_L$ the unique, sized $R$, leaf and $\brac{U_{PORTS}}$ the ports on that leaf.
The higher level algorithm considering the possible valid combinations of $Q,D$ and $R$, for 2-level and 3-level fat-trees is provided in Algorithm~\ref{alg:alg2}.

\ST{Extension for over-subscribed fat trees}
In order to reduce the network equipment cost, some cloud vendors use over-subscribed fat-trees, also known as \emph{slimmed fat-trees}~\cite{rodriguez_oblivious_2009}.
In an over-subscribed fat-tree, the number of uplinks is smaller than the number of downlinks in the switches, contrarily to the full bisectional bandwidth fat-tree, where they are equal. (We assume equal-bandwidth links).
In such trees, we denote $O_i$ the ratio between the two total number of links: those connecting switches at level $i$ to the previous level $i-1$, and those connecting to the next level $i+1$. By this definition for XGFT:
\begin{equation}
O_i=\frac{m_i}{w_{i+1}}
\end{equation}
We describe here how to provide LaaS for over-subscribed fat-trees, without requiring hardware-assisted accurate TDMA link sharing.
For simplicity we do not support tenant selection of their requested bandwidth.
Since we allow no link-sharing between tenants, and we have no preference between tenants, a tenant placed across a level $i$ of the tree has at least $O_i$ permutation flows shared on each link.
So for crossing level $i$ we only require $S$ common switches at level $i+1$:
\begin{equation}\label{eq:slimming}
S=\abs{S^D}=\ceil{\frac{D}{\ceil{O_i}}}
\end{equation}

Clearly, a selection of $D$ such that it is not divisible by $\ceil{O_i}$ reduces the cluster utilization, so the order by which we search for sub-trees should reflect that priority.
The changes to Algorithm~\ref{alg:alg1} are a new function argument $S$ which defines the number of spines required, and its usage in line 7: \textit{$\textbf{if}~r = S~\textbf{then}$}.
The changes to Algorithm~\ref{alg:alg2} involve adding an $S$ of~\eq{eq:slimming} to the calls of $FLAP$ and also adding an external loop around the \textit{\textbf{for}} statements in lines $11-19$ and $23-31$ to try $D$ values divisible by $\ceil{O_i}$ first.
}

\section{Evaluation}\label{sec:evaluation}
Our evaluation is reported in three sub-sections.
The first deals with the resulting \emph{cloud utilization} when applying LaaS conditions.
It shows that our \emph{LaaS} algorithm reaches a reasonable cloud utilization, within about $10\%$ of bare-metal allocation.
The second part describes the system implementation on top of OpenStack, and the third part shows how the LaaS architecture improves the performance of a tenant in the presence of other tenants by completely isolating the tenants from each other.

\subsection{Evaluation of Cloud Utilization}
\T{Cloud utilization.} We want to study whether our LaaS network isolation constraints significantly reduce the number of hosts that can be allocated to tenants. We define the \emph{cloud utilization} as the average percentage of allocated hosts in steady state. Assuming that tenants pay a fee proportional to the number of used hosts and the time used, the cloud utilization is a direct measure of the revenue of the cloud provider.

\T{Scheduling simulator.} To evaluate the different heuristics on large-scale clouds, we developed a scheduling simulator that runs many tenant requests over a user-defined topology.
The simulator is configured to run any of the above algorithms for host and link allocation.
This algorithm may succeed and place the tenant, or fail. We use a strict FIFO scheduling, i.e. when a tenant fails, it blocks the entire queue of upcoming tenants.
Note that this blocking assumption forms an extremely conservative approach in terms of cloud utilization.
In practice, clouds would typically not allow a single tenant to block the entire queue and use resource reservation with back-filling techniques to overcome such cases.
Since smaller tenants are easier to place, for any tenant size distribution, not letting smaller tenants bypass those waiting means that we fill fewer tenants into the cloud. Thus, the result should be regarded as an intuitive lower-bound for a real-life cloud utilization.

\begin{figure}[]
\centering
\includegraphics[width= 0.9 \columnwidth]{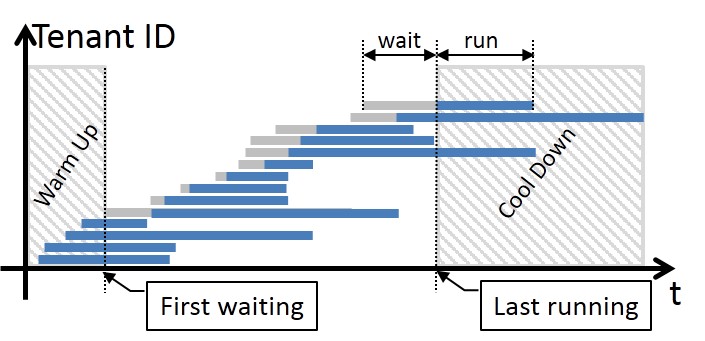}
\caption{Utilization is measured after the first tenant cannot enter the cloud and before the cloud starts draining out of tenants.}
\label{fig:effciency}
\end{figure}

\ORM{
All host placement heuristics need to eventually select a possible placement from many valid ones.
Our heuristic first try larger $D$ values which spreads the tenant over lower number of sub-trees.
Once the $D$ is decided we follow a bin-packing approach that tries to use the most-used leaf switches first (i.e., those with the least number of available hosts). This is known to minimize fragmentation~\cite{grandl2014multi}. We also tested other placement-order policies, like preferring a larger spread of the tenants, but they produced inferior results.
}

\T{Simulation settings.} We simulate the scheduler with LaaS algorithm on the largest full-bisectional-bandwidth 3-level fat-tree network that can be built with 36-port switches, i.e. a cloud of 11,662 hosts.
The evaluation uses a randomized sequence of 10,000 tenant requests. A random run-time in the range of 20 to 3,000 time units is assigned to each tenant.
The variation of run-time makes scheduling harder as it increases fragmentation.

We evaluate 4 distribution types for the number of hosts requested by incoming tenants.
First, we randomly generate sizes according to a job size distribution extracted from the Julich JUROPA job scheduler traces.
These previously-unpublished traces represent 1.5 years of activity (Jan. 2010 - June 2011) of Julich JURUPA,
a large high-performance scientific-computing cloud.
Second, we use a truncated exponential distribution of variable average $x$.
It is  truncated between 1 and the cluster size.
Then, we evaluate a truncated Gaussian distribution of average parameter $x$ and standard deviation parameter $\frac{x}{5}$.
Last, we evaluated a uniform distribution of tenant sizes with a variable average $x$ and range of $\left[0.2x,1.8x\right]$.


As a baseline algorithm, we implement an \emph{Unconstrained} placement approach that simply allocates unused hosts to the request, as in bare-metal allocation.
Note that some requests may still fail if the tenant requests more hosts than the number of currently-free cloud hosts. We compare this baseline to the \emph{Simple} and \emph{LaaS} algorithms, as described in Section~\ref{sec:analysis}.

\T{Simulation results.}
\fig{fig:julich_cdf} illustrates the Cumulative Distribution Function (CDF) of the tenant sizes (in number of hosts) collected from the Julich JUROPA cluster. 
The CDF shows peaks for numbers of hosts that are powers of 2 (1, 2, 4, 8, 16, and 32).
We further generated 10,000 tenants with this job-size probability distribution, and the same random run-time distribution as above (instead of the original run-times, since they resulted in a low load, and therefore in an easy allocation).
\fig{fig:julich_util} shows the tenant allocation results: the cost of our \emph{LaaS} allocation versus the \emph{Unconstrained} bare-metal provisioning is about 10\% of cloud utilization ($88\%$ vs. $98\%$).
\begin{figure}[t]
\centering
\subfigure[] {
\includegraphics[width=0.42 \columnwidth]{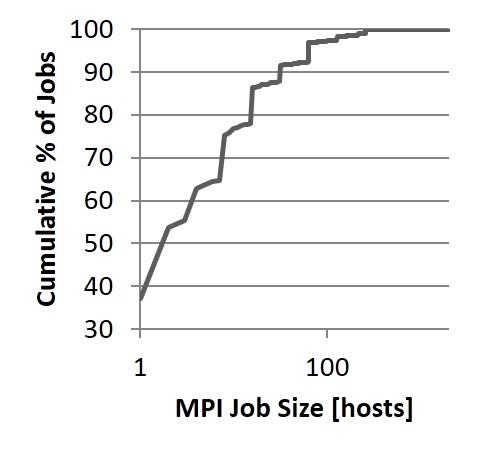}
\label{fig:julich_cdf}}
\subfigure[] {
\includegraphics[width=0.45 \columnwidth]{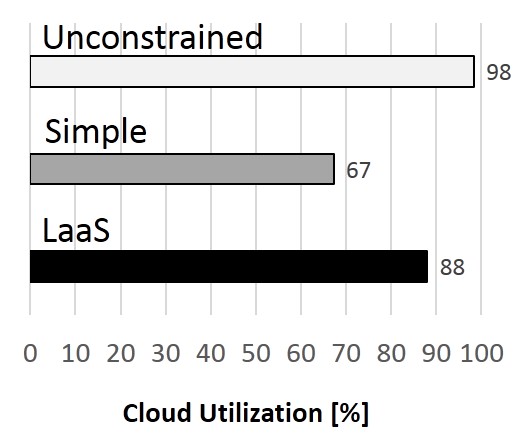}
\label{fig:julich_util}}
\caption{(a) Measured job-size Cumulative Distribution Function (CDF) for the Julich JUROPA scientific-computing cloud. (b) Resulting cloud utilization. 
\emph{LaaS} achieves 88\%.}
\label{fig:julich_cdf_util} \vspace{-0.3cm}
\end{figure}

To further test the sensitivity of our algorithm to the tenant sizes, we use a truncated exponential distribution for tenant host sizes and modify the exponential parameter $x$.
The distribution of the JUROPA tenant sizes is similar to such a truncated exponential distribution. \fig{fig:exp_pnr} illustrates the cloud utilization for \emph{Unconstrained}, \emph{Simple}, and \emph{LaaS}, is plotted as a function of the exponential parameter $x$, which is close to the average tenant host size due to the truncation.
The \emph{Unconstrained} line shows how the utilization degrades with the job size, even without any network isolation.
This is an expected behavior of bin packing. As the job size grows, so does the probability for more nodes to be left unassigned when the cloud is almost full.
The utilization of our \emph{LaaS} algorithm stays steadily at about 10\% less than the \emph{Unconstrained} algorithm.
Finally, \emph{Simple} has the lowest cloud utilization for the entire tenant size range.
Note that it is {less} steady, since its utilization is more closely tied to the sizes of the leaves and sub-trees. Once the tenant size crosses the leaf size (18 in our case), it is rounded up to a multiple of that number. Likewise, once it crosses the size of a complete sub-tree (324 hosts), it is rounded up to the nearest multiple of that number.
These results show that our LaaS algorithm provides an efficient solution for avoiding tenant variability, as its cost is only about 10\% for a wide range of tenant sizes.
\begin{figure}[]
\centering
\includegraphics[width= 0.98 \columnwidth]{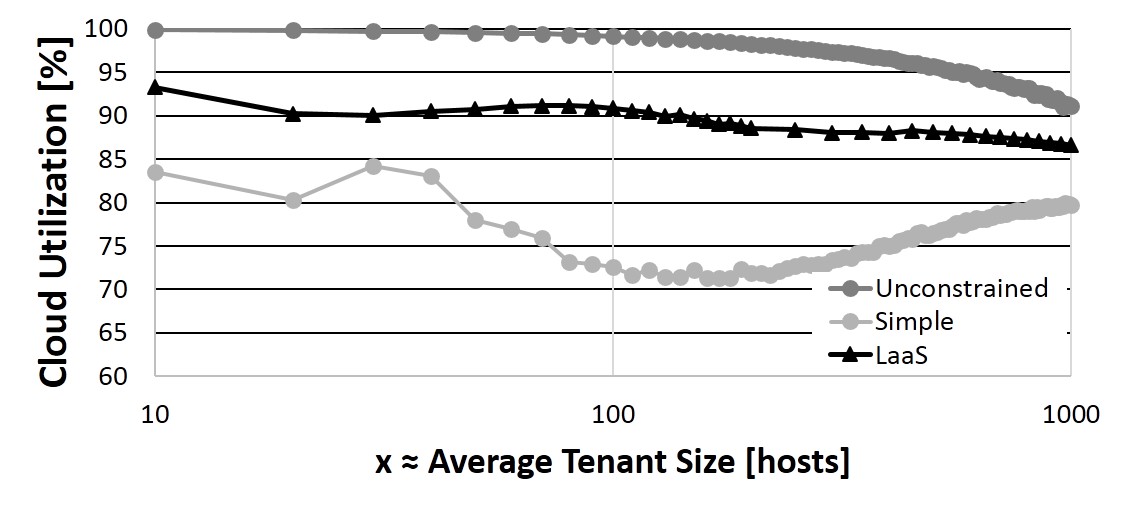}
\caption{Cloud utilization for a truncated exponential distribution of tenant host sizes in a cloud of 11,662 hosts.}
\label{fig:exp_pnr}
\end{figure}

\fig{fig:gauss} illustrates the cloud utilization for the truncated Gaussian distribution. This distribution provides a harder test for the allocation algorithm, since tenant sizes are made similar, and they may be just beyond the above-mentioned thresholds of a leaf size (18 hosts) or a sub-tree size (324 hosts).
These thresholds are where \emph{LaaS} and the \emph{Simple} are less efficient when compared to \emph{Unconstrained}.

\ORM{
\begin{figure}[t]
\centering
\includegraphics[width= 0.98 \columnwidth]{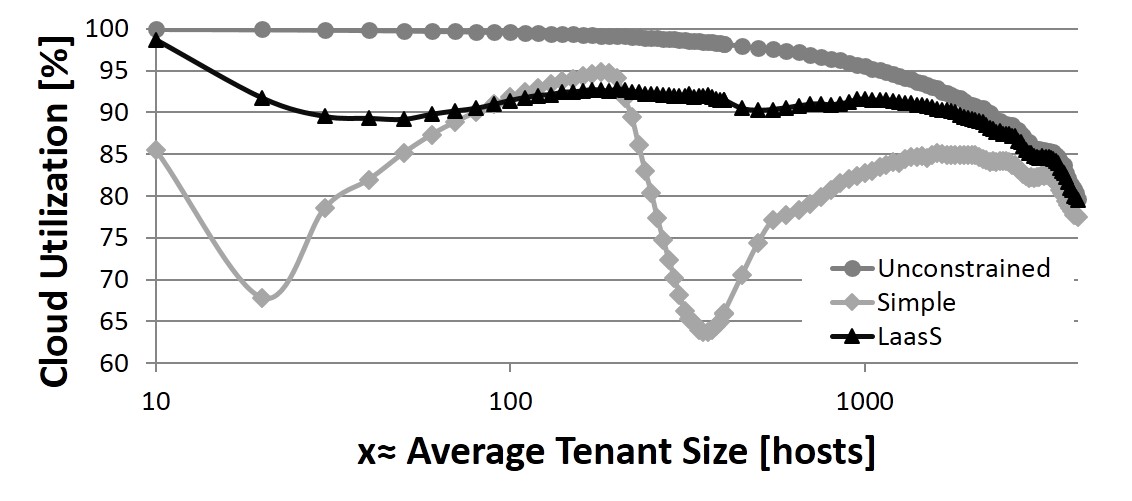}
\caption{Cloud utilization for a truncated Gaussian distribution $\mathcal{N}(x,x/5)$ of tenant host sizes in a cloud of 11,662 hosts.}
\label{fig:gauss_util} \vspace{-0.3cm}
\end{figure}
}

\begin{figure}[t]
\centering
\includegraphics[width= \columnwidth]{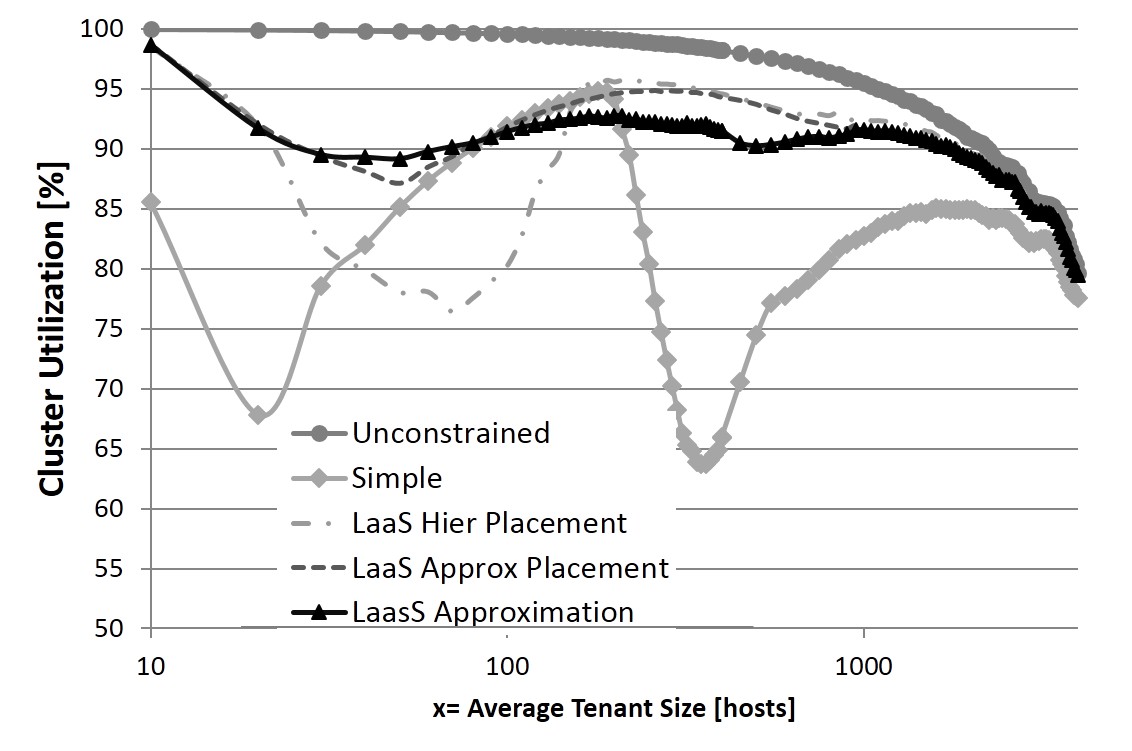}
\caption{Cloud utilization for a truncated Gaussian distribution $\mathcal{N}(x,x/5)$ of tenant host sizes in a cloud of 11,662 hosts.}
\label{fig:gauss}
\end{figure}

\emph{Simple} suffers from a particularly large fluctuation in utilization.
\emph{LaaS} is more stable over the entire range, with about 90\% utilization. There are a few points where the \emph{Simple} heuristic provides a better utilization than \emph{LaaS}. But, note that utilization stability is key to cloud vendors, since changing the allocation algorithm dynamically would require predicting the future size distribution, and thus may produce worse results when the distribution does not behave as expected.

~\fig{fig:ifb_stddevs} plots the \emph{LaaS Approximation} utilization for different spreads of tenant sizes around the average. A standard deviation of avg/5, avg/10 and 0 are shown.
The zero deviation curve exhibits the expected saw-tooth shape that is caused by the fact it is possible to get 100\% utilization when the tenant size is a divisor of the number of nodes.
As the deviation of the tenant sizes grows so does the smoothness of the curve. This is common to all scheduling algorithms behavior providing the peaks and valleys around the job sizes crossing the singe leaf or sub-tree size.
\begin{figure}[]
\centering
\includegraphics[width= \columnwidth]{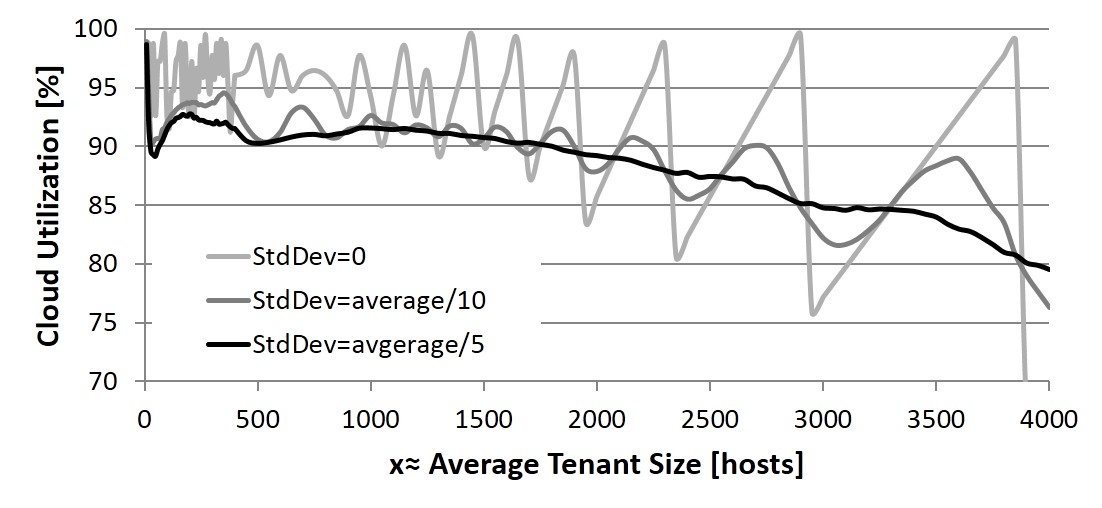}
\caption{Cloud utilization for a truncated Gaussian distribution $\mathcal{N}(x,\alpha)$ of tenant host sizes in a cloud of 11,662 hosts, for $\alpha \in \brac{0,x/10,x/5}$.}
\label{fig:ifb_stddevs}
\end{figure}

Utilization obtained for the uniform distribution of tenant size is presented in~\fig{fig:uniform_pnr}.
As can be seen there is clear advantage to the LaaS Placement heuristic that maintains utilization of about ~90\%.
\begin{figure}[t]
\centering
\includegraphics[width= \columnwidth]{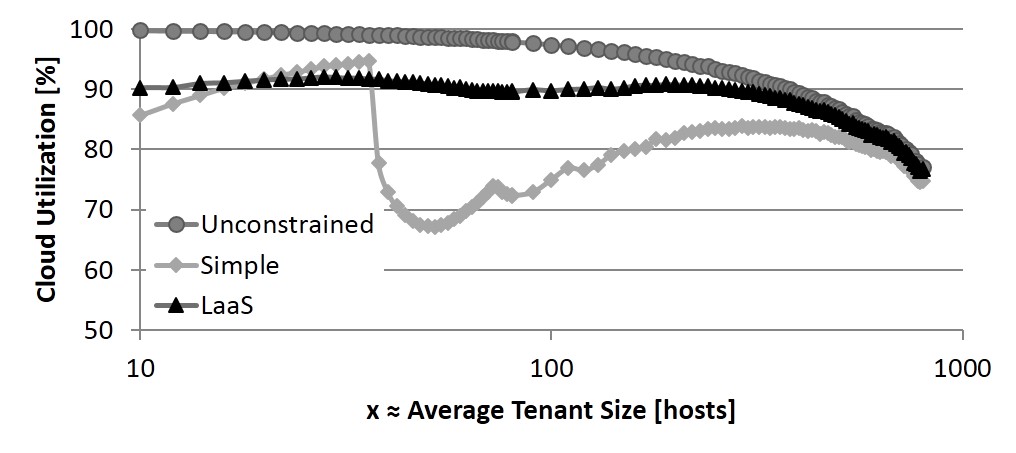}
\caption{Cloud utilization vs. average tenant size for 10,000 requests with Uniform(0.2x,1.8x) size distribution.}
\label{fig:uniform_pnr}
\end{figure}

LaaS also provides the opportunity to turn off unused links that are not allocated to tenants.
\fig{fig:off_links} provides the percentage of links that could be turned off, for the LaaS scheduling of the Julich distribution of tenant sizes.
As can be observed, the average percentage of links that could be turned off is linear with the cloud utilization.
As the utilization decreases the number of unused links grows accordingly and the network power can be linearly reduced.
\begin{figure}[!t]
\centering
\includegraphics[width= \columnwidth]{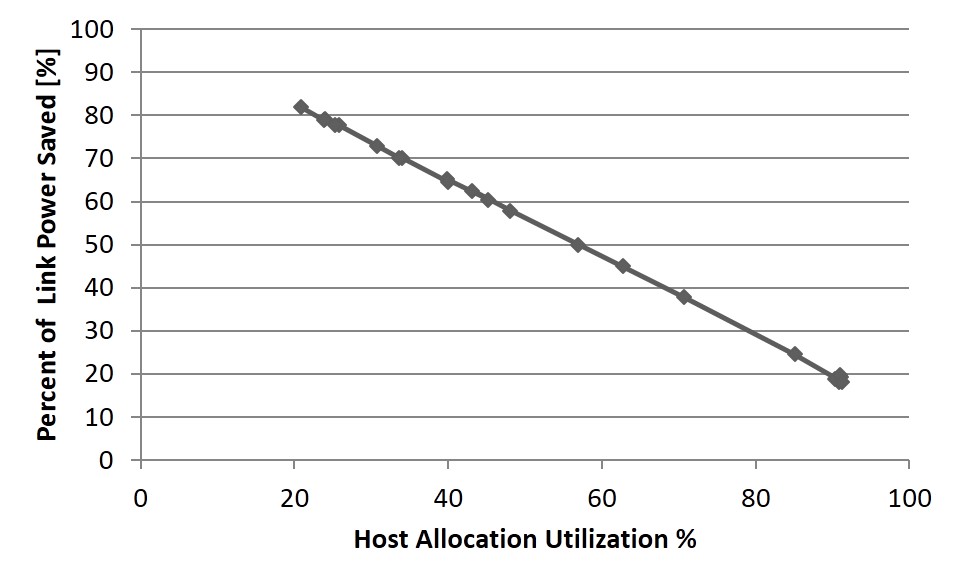}
\caption{Percentage of links that can be turned off in the 3-level fat-tree as a function of the cloud utilization.}
\label{fig:off_links}
\end{figure}

\subsection{System Implementation}
We implemented the LaaS architecture by extending the \emph{OpenStack Nova scheduler} with a new service that first runs the LaaS host and link allocation algorithm, and then translates the resulting allocation to an SDN controller that enforces the link isolation via routing assignments.

\T{Host and link allocation.} The integration of the \emph{LaaS} algorithm was done on top of OpenStack (Icecube release), utilizing filter type: \emph{AggregateMultiTenancyIsolation}.
{This filter allows limiting tenant placement to a group of hosts declared as an ``aggregate'', which is allocated to the specific tenant-id.}
Our automation, provided as a standalone service on top of OpenStack's \emph{nova} controller, obtains new tenant requests, and then calls the \emph{LaaS} allocation algorithm.
If the allocation succeeds, we invoke the 
command to create a new aggregate that is further marked by the tenant-id. 
The allocated hosts are then added to the aggregate. 
The filter guarantees that a new host request, conducted by a user that belongs to a specific tenant, is mapped to a host that belongs to the tenant aggregate.

\T{Network controller.} We further implement a method to provide the link allocation to the InfiniBand SDN controller~\cite{ofa_opensm_git_2002}, which allows it to enforce the isolation by changing routing.
The controller supports defining sub-topologies, by providing a file with a list of the switch ports and hosts that form each sub-topology.
Then each sub-topology may have its own policy file that determines how it is routed.
We ran the SDN controller over the simulated network of 1,728 hosts, as well as over our 32-host experimental cluster.

\T{Run-time.} 
The LaaS Approximation scans through all possible placements for valid link allocation.
This involves evaluating all possible valid combinations of $R$ and $Q$ values.
~\fig{fig:alloc_runtime} presents the 
average run-time per tenant request for placing tenants on 11,664 nodes cluster providing a truncated exponential tenant size distribution. Run time was measured on an Intel\textsuperscript{\textregistered} Xeon\textsuperscript{\textregistered} CPU X5670 @ 2.93GHz.
The peak in run-time of about $5$~msec appears just below the average tenant size of 324, which is the exact point where our algorithm first scans all possible placements under a single sub-tree and continues with multiple sub-tree placement.
{
\begin{figure}[t]
\centering
\includegraphics[width= 0.9 \columnwidth]{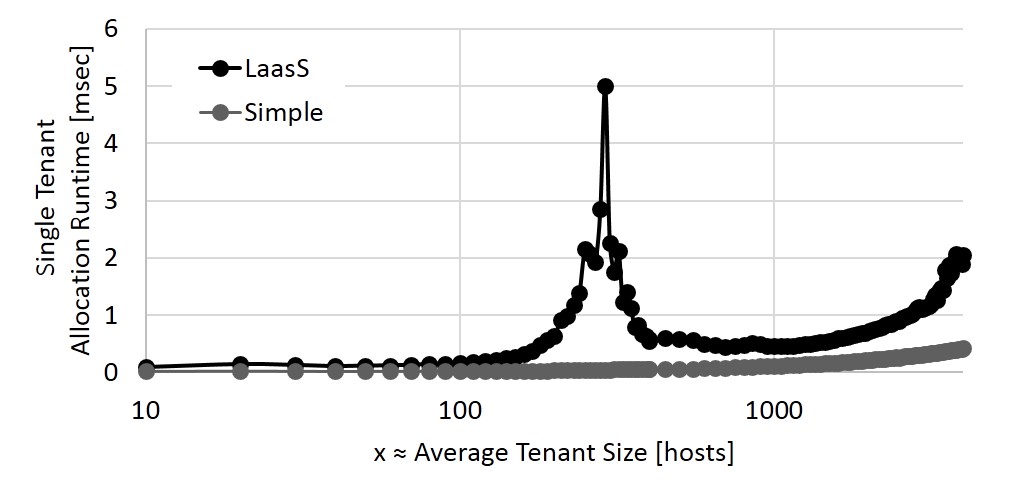}
\caption{Average run-time of single tenant allocation versus average tenant size.}
\label{fig:alloc_runtime}
\end{figure}
}

\subsection{Evaluation of Tenant Performance}
{Since LaaS guarantees tenant isolation, tenant performance should be independent of the number of other tenants that run on the same network.
}
To demonstrate LaaS tenant isolation, we simulate a large cluster using a well known InfiniBand flit level simulator used by~\cite{domke_deadlock-free_2011,zahavi_fat-tree_2012,gran_exploring_2012}.

\fig{fig:lass_32_jobs} presents the relative performance of single and multiple tenants running Stencil scientific-computing applications on a cloud of 1,728 hosts, under either \emph{Unconstrained} or \emph{LaaS}, normalized by the performance of a single tenant placed without constraints.
The figure illustrates many effects. First, the performance of a single tenant with \emph{Unconstrained} significantly degrades when other tenants are active, e.g. to 45\% with 32-KB message sizes. This is because the bare-metal allocation of \emph{Unconstrained} does not provide link isolation. Second, 
under our \emph{LaaS} algorithm, \emph{the single-tenant performance is not impacted when the other tenants become active} (the third and fourth sets of columns look identical). This was the key goal of this work. \emph{LaaS} prevents any inter-tenant traffic contention. Finally, we can observe an additional surprising effect (first vs. third sets of columns):  the tenant performance is slightly improved for small messages under \emph{LaaS} versus the \emph{Unconstrained} allocation. The reason is that \emph{LaaS} does not accept tenants unless it can place them with no contention, and therefore the resulting placement tends to be tighter, thus improving the run-time performance with small message sizes when the synchronization time of the tasks is not negligible. The lower network diameter of \emph{LaaS} improves the synchronization time, which is latency-dominated.

\begin{figure}[]
\centering
\includegraphics[width= 0.9 \columnwidth]{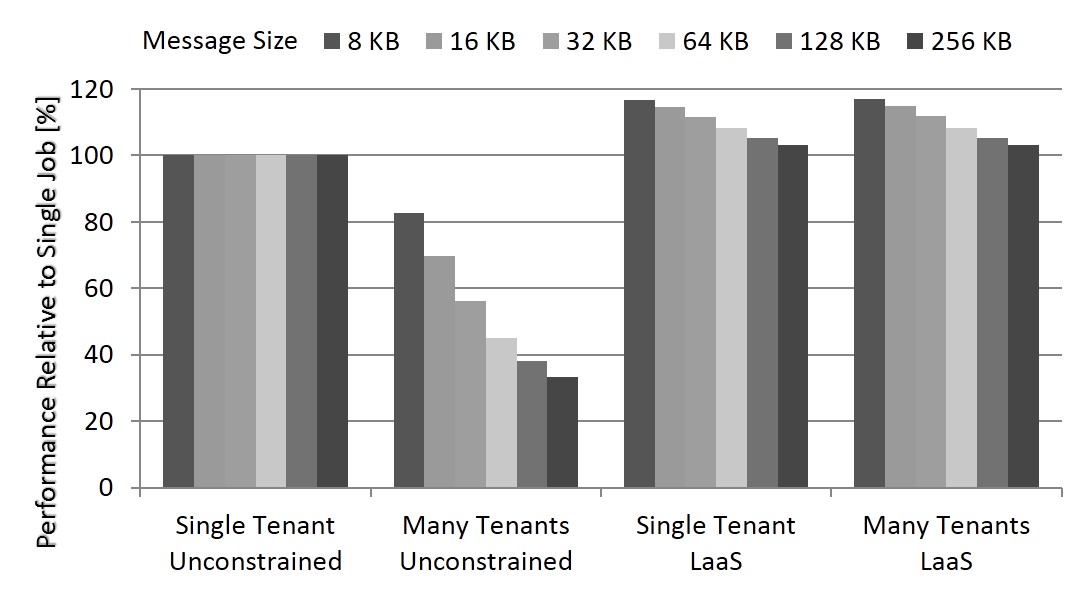}
\caption{Simulated relative performance for tenants running Stencil scientific-computing applications on a cloud of 1,728 hosts, either alone or as 32 concurrent tenants. While tenant performance degrades when placed unconstrained (without link isolation), the performance of single and multiple tenants with LaaS appears identical, fulfilling the promise of LaaS.}
\label{fig:lass_32_jobs} \vspace{-0.3cm}
\end{figure}
\ORM{
The simulated cloud evaluation uses the routing obtained from OpenSM as the forwarding rules in the dynamic flit level InfiniBand network simulator.
Then traffic of MPI programs is then injected into the network and their run-time is measured when running single or with other tenants on the cloud.
The results of this experiment are provided in ~\fig{fig:sim_isolation} which shows the MPI jobs performance is completely independent of each other.
\begin{figure}[]
\centering
\includegraphics[width= \columnwidth]{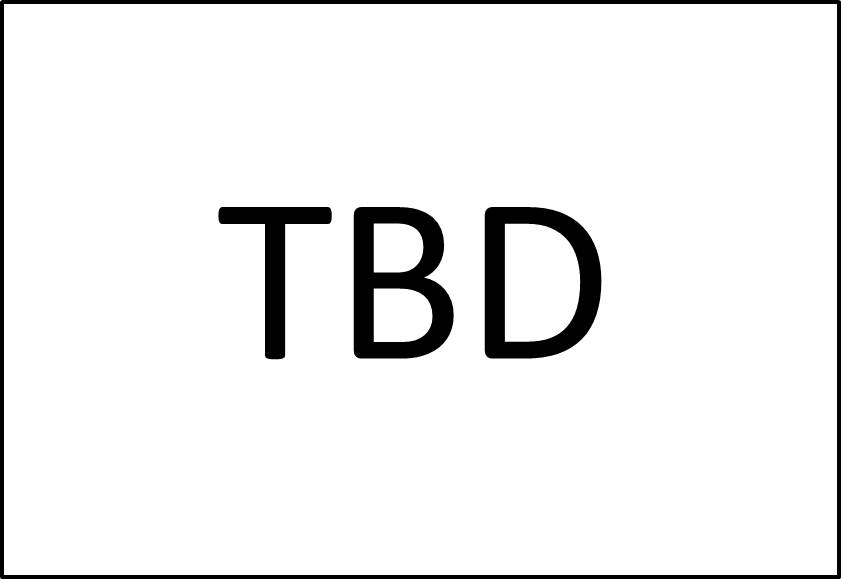}
\caption{Isolated 2,9 and 32 MPI jobs simulated on 1,728 hosts cloud. The simulation shows the MPI jobs performance is completely independent of each other.}
\label{fig:sim_isolation}
\end{figure}
For the test bench cloud testing we provide the tenant placement to an OpenSM controlling our 32 nodes cloud and then gradually invoke the 4 MPI jobs and measure their iteration run-time.
The measured run-time shown in ~\fig{fig:run_isolation} proves the isolation works and job run-time is independent of other jobs.

\begin{figure}[]
\centering
\includegraphics[width= \columnwidth]{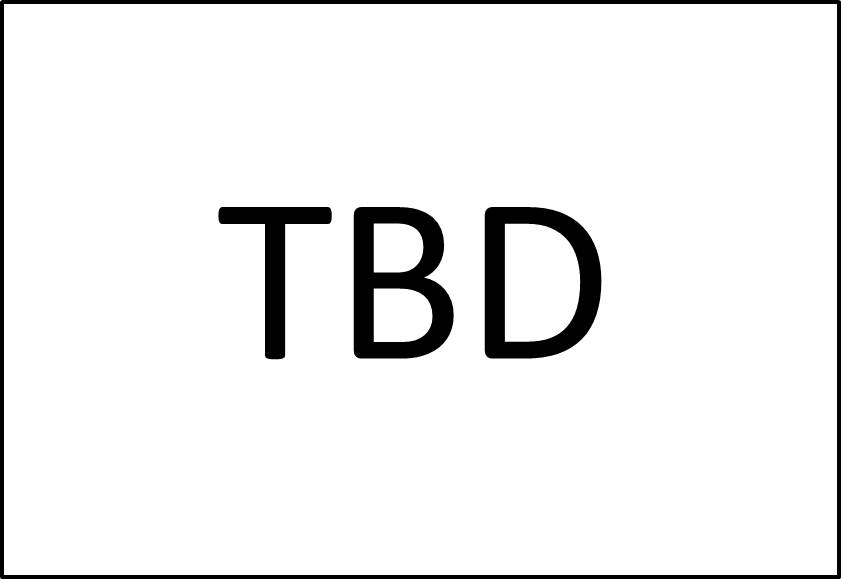}
\caption{Isolated 4 MPI jobs run on 32 hosts InfiniBand 10Gbps cloud. The iteration run-time is showing no interdependence between the jobs.}
\label{fig:run_isolation}
\end{figure}
\begin{itemize}
\item 3. Show MPI simulations results for 2,9,32 jobs placed with isolation
\end{itemize}
} 

\section{Discussion}\label{sec:discussion}

\T{Recursive LaaS.}
When talking to industry vendors, they pointed out simple extensions that would easily generalize the use of LaaS. First, LaaS could be applied recursively, by having each tenant application or each sub-tenant reserve its own chunk of the cloud within the tenant's chunk of the cloud. Second, LaaS could also be applied in private clouds, with cloud chunks being reserved by applications instead of tenants. Third, shared-cloud vendors could easily restrict LaaS to a subset of their cloud, while keeping the remainder of their cloud as it is today. This can be done by reserving large portions of the topology to a \textit{virtual} tenant that is shared between many real tenants. Pre-allocation and modification of that sub-topology is already supported by our code. As a result, LaaS offers a smooth and gradual transition to better service guarantees, enabling cloud vendors to start only with the tenant owners who are most ready to pay for it.

\T{Off-the-shelf LaaS.} LaaS is implementable today with no extra hardware cost in existing switches and no host changes.
The algorithm requires only a moderate software change in the allocation scheme, which we provide as open source.
It also relies on an isolated-routing feature of the SDN controller, which is already available in InfiniBand and could be implemented in Ethernet SDN controllers like OpenDaylight.

\T{Proportional network power.} LaaS eases the use of an elastic network link power that would be made proportional to cloud utilization~\cite{abts_energy_2010}. This is because it explicitly mentions which links and switches are to be used, and therefore can turn off other links and switches. In other approaches the control has to happen as a result of traffic load change and thus is not realistic for common switch hardware for which the turn-ON time is much larger than a microsecond.

\T{Heterogeneous LaaS.}
Host allocation in heterogeneous clouds involves allowing tenants to express their required host features in terms of CPU, memory, disk and available accelerators.
On such systems, the host allocation algorithm should allow the provider to trade off the acceptance of a new tenant versus the cost of the available hosts, which may be higher as their capabilities may exceed the user needs.
Our LaaS algorithm could support these requirements.
\emph{Although this requirement complicates the allocation algorithm, it is feasible to support it in LaaS.}
First, it should use the host costs to order the search. Second, it should try all the possible divisors and select the one with best accumulated cost.
A trade-off between the resulting fragmentation and the cost difference could extend it.

\T{LaaS with VMs.} LaaS could easily support multiple tenants running as virtual machines (VMs) on the same host, assuming accurate packet pacing and burst control is provided by hosts and switches. LaaS could then treat each link as a set of isolated links and assign them to different tenants. This includes the links leaving the host.


\T{Non-FIFO tenant scheduling.} We conservatively evaluated our \emph{LaaS} allocation algorithm assuming FIFO scheduling of incoming tenants. To improve the cloud utilization, we could equally rely on a non-FIFO policy, e.g. by using back-filling, reservations, or a jointly-optimal allocation of multiple tenants~\cite{pascual_job_2009}.
\ORM{
\T{Energy Proportionality} is another benefit of LaaS.
Since links are explicitly allocated, those links left un-allocated can be proactively turned off and save power, as opposed to only partially throttled down~\cite{abts_energy_2010} when traffic is load balanced over the network.
We show that with LaaS the number of used links is proportional to the cloud utilization.
}

\T{Fault Tolerance.} When a link is down before being allocated it is easy to avoid allocating it to new tenants.
However, if a link was already allocated to a tenant, it is not always possible to provide an alternative link without breaking the current operation of the tenant.
Similarly to losing a link on the private cloud, the tenant will see some degradation until the link is fixed or the forwarding plane is adapted.

\section{Conclusions} \label{sec:conc}
In this paper, we demonstrated that the interference with other tenants causes a performance degradation in cloud applications that may exceed 65\%.
We introduced LaaS (Links as a Service), a novel cloud allocation and routing technology that provides each tenant with the same bandwidth as in its own private data center. We showed that LaaS completely eliminates the application performance degradation. We further explained how LaaS can be used in clouds today without any change of hardware, and showed how it can rely on open-source software code that we contributed.
Finally, we also used previously-unpublished tenant-size statistics of a large scientific-computing cloud, obtained over a long period of time, to construct a random workload that illustrates how isolation is possible at the cost of some 10\% cloud utilization loss.

%
%




\bibliographystyle{IEEEtran}
\bibliography{laas} 
%
%
%

\begin{appendices}     
\section{LaaS Software Release 1.0}\label{sec:source_code}
The software is provided in~\cite{dropbox_laas_2015} under the directory laas\_1.0/ as well as in a single archive file: laas\_1.0.tgz.
In this section, we provide all the information required to get the LaaS service installed, and instructions to
run a demonstration of the service.
We also provide the simulation setup used for obtaining the cluster utilization, run-time and correctness.

The simulator and a service of LaaS are coded in Python and are built on top of the core algorithm coded in C++.
At the heart of the package is the LaaS algorithm coded in \emph{isol.cc}. It is using facilities specific to 3-level fat-trees provided in \emph{ft3.cc} and port-mask utility class in \emph{portmask.cc} used for tracking availability of links.
The \emph{laas.cc} implements the service API provided in \emph{laas.h} and exposed in Python using SWIG which uses the declarations in \emph{laas.i}.
We provide a scheduling simulator, to obtain cluster utility, in \emph{sim.py} and a tenant allocation service in \emph{laas\_service.py}.

The LaaS service provides a RESTful interface and serves tenant requests~\cite{richardson_restful_2007}.
It outputs OpenStack command files required to control tenant host placement and also
provides SDN configuration files to enforce isolation via packet routing/forwarding.

The scheduling simulator takes a CSV file with tenant requests (id, size and arrival time) and process them
in a FIFO manner.

\subsection{License}
This software is provided with a choice of GPLv2 or BSD license and published on our website.

\subsection{Content}
The following sub-directories are included in this release:

\begin{itemize}
\item \emph{src} - The core algorithm c++ and python service/simulator
\item \emph{bin} - Random tenants generator and isol.log checker
\item \emph{examples} - A set of files used by the demo below
\end{itemize}

Out of the entire set of source files, the one most interesting for integration is
lass.h which provides the API exposed in Python.

\subsection{Software dependency}
\begin{itemize}
\item Any Linux environment, for example Ubuntu 12.04
\item SWIG Version 2.0.11
\item Python 2.7
\item Python 2.7 Flask 0.7
\item Python 2.7 Flask Restful 0.3.1
\end{itemize}

The Perl code (for utilities only) depends on:
\begin{itemize}
\item Perl v5.18.2
\item Perl Math::Random 0.71
\item Perl Math::Round 0.07
\end{itemize}

\subsection{Installation}
{\scriptsize\begin{verbatim}
tar xvfz laas_1.0.tgz
cd laas_1.0/src
make
\end{verbatim}}

\subsection{Running LaaS Service}
\begin{enumerate}

\item \emph{Choose your cluster topology:}\\
   For ease of review we choose a small 2 level fat tree.
   The example topology is XGFT(2; 4,8; 1,4). Due to limitation of the current implementation
   we represent it as if it were a 3 level fat tree with one top switch: XGFT(3; 4,8,1; 1,4,1)
   The data needed to run a larger topology is also included in the examples directory.

\item \emph{Prepare name mapping file:}\\
   The LaaS engine eventually needs to configure OpenStack and an SDN controller that rely on
   physical naming and port numbering and not on general fat-tree indexing. A file that
   provides mapping of the tree level, index within the tree and port indexing to the actual cluster hardware
   is thus required. For this example topology we provide the mapping file: examples/pgft\_m4\_8\_w1\_4.csv.

The first line hints at the content of
each column:
{\scriptsize\begin{verbatim}
# lvl,swIdx,name,UP,upPorts,DN,dnPorts
\end{verbatim}}
The example line below describes a host,
providing its level is 0 and index is 10,
its name is comp-11 and it has a single
UP port, number 1, connecting to L1 switch (on level 1).
{\scriptsize\begin{verbatim}
0,10,comp-11,UP,1,,,,,,,,,,,,,,,,,,,,,,,,
\end{verbatim}}
An example L1 switch line is provided below.
See this is the 4\textsuperscript{th} switch in L1, its name as
recognized by the SDN controller is SW\_L1\_3
and its ports 5-8 are connecting to hosts:
{\scriptsize\begin{verbatim}
1,3,SW_L1_3,UP,1,2,3,4,DN,5,6,7,8
\end{verbatim}}
Note: The file does not include any mapping for the
non existant level 3 switches.

\item \emph{Start the service:}\\
Once started the LaaS service reports its address and port.
The Restful API is up and any change in tenant status will result in updates in the
OSCfg/ and SDNCfg/ directories.
{\scriptsize
\begin{verbatim}
|$ python ./src/laas_service.py  -m 4,8,1 -w 1,4,1 \
    -n examples/pgft_m4_8_w1_4.csv	
|-I- Defined 64 up ports and 64 down port mappings				
|  										
|* Running on http://127.0.0.1:12345/						
|* Restarting with reloader							
|-I- Defined 64 up ports and 64 down port mappings
\end{verbatim}
}

\item \emph{Run a demo:}\\
We provide here an example sequence of calls to the service.
After each step we discuss the results and the created files if any.

\subsubsection{List tenants:}
{\scriptsize\begin{verbatim}
| $ curl http://localhost:12345/tenants
| {}
\end{verbatim}}
As expeted it returns an empty list
\subsubsection{Create a tenant of 10 nodes:}
(Expecting it will span 2.5 leafs.)
{\scriptsize\begin{verbatim}
 $ curl http://localhost:12345/tenants -d "id=4" -d "n=10" -X POST
| {
|	"N": 10,
| 	"hosts": 10,
|  	"l1Ports": 10,
|	"l2Ports": 0
| }
\end{verbatim}}
See how the tenant-id may be any number for which there is no pre-existing tenant in the system.
Let's inspect the created files. First see the new file in the OSCfg:
{\scriptsize\begin{verbatim}
cmd-1.log:
| #!/bin/bash
| #
| # Adding tenant 4 to OpenStack
| #
| echo Adding tenant 4 to OpenStack > OSCfg/cmd-1.log
| keystone tenant-create --name laas-tenant-4 \
|   --description "LaaS Tenant 4" >> OSCfg/cmd-1.log
| tenantId=`keystone tenant-get laas-tenant-4 | \
|   awk '/ id /{print $4}'` >> OSCfg/cmd-1.log
| nova aggregate-create laas-aggr-4 >> OSCfg/cmd-1.log
| nova aggregate-set-metadata laas-aggr-4 \
|   filter_tenant_id=$tenantId >> OSCfg/cmd-1.log
| nova aggregate-add-host laas-aggr-4 comp-1 >> cmd-1.log
| nova aggregate-add-host laas-aggr-4 comp-2 >> cmd-1.log
| nova aggregate-add-host laas-aggr-4 comp-3 >> cmd-1.log
| nova aggregate-add-host laas-aggr-4 comp-4 >> cmd-1.log
| nova aggregate-add-host laas-aggr-4 comp-5 >> cmd-1.log
| nova aggregate-add-host laas-aggr-4 comp-6 >> cmd-1.log
| nova aggregate-add-host laas-aggr-4 comp-7 >> cmd-1.log
| nova aggregate-add-host laas-aggr-4 comp-8 >> cmd-1.log
| nova aggregate-add-host laas-aggr-4 comp-9 >> cmd-1.log
| nova aggregate-add-host laas-aggr-4 comp-10 >> cmd-1.log
\end{verbatim}}
Similarly, the SDNCfg/ directory now holds a full set of configuration files
required for OpenSM to configure the network. We will not go through the
full description of these files but focus on the groups.conf.
This file now holds the definition of the hosts and switch ports used by the first tenant:
{\scriptsize\begin{verbatim}
| port-group
| name: T4-hcas
| obj_list:
|    name=comp-1/U1:P1
|    name=comp-2/U1:P1
|    name=comp-3/U1:P1
|    name=comp-4/U1:P1
|    name=comp-5/U1:P1
|    name=comp-6/U1:P1
|    name=comp-7/U1:P1
|    name=comp-8/U1:P1
|    name=comp-9/U1:P1
|    name=comp-10/U1:P1;
| end-port-group
|
| port-group
| name: T4-switches
| obj_list:
|    name=SW_L1_2/U1 pmask=0x6
|    name=SW_L1_0/U1 pmask=0x1e
|    name=SW_L1_1/U1 pmask=0x1e;
| end-port-group
\end{verbatim}}
\subsubsection{To fill in the network we create another 10-node tenant:}
{\scriptsize\begin{verbatim}
| $ curl http://localhost:12345/tenants -d "id=1" -d "n=10" -X POST
| {
|	"N": 10,
| 	"hosts": 10,
|  	"l1Ports": 10,
|	"l2Ports": 0
| }
\end{verbatim}}
\subsubsection{List again the tenants:}
{\scriptsize\begin{verbatim}
| $ curl http://localhost:12345/tenants
| {
|     "1": {
|         "N": 10,
|         "hosts": 10,
|         "l1Ports": 10,
|         "l2Ports": 0
|     },
|     "4": {
|         "N": 10,
|         "hosts": 10,
|         "l1Ports": 10,
|         "l2Ports": 0
|     }
| }
\end{verbatim}}
\subsubsection{Get the allocated hosts and links for a specific tenant:}
{\scriptsize\begin{verbatim}
| $ curl http://localhost:12345/tenants/1/hosts
| [
|     "comp-11",
|     "comp-12",
|     "comp-13",
|     "comp-14",
|     "comp-15",
|     "comp-16",
|     "comp-17",
|     "comp-18",
|     "comp-19",
|     "comp-20"
| ]
\end{verbatim}}
As expected the four spines are going to be used (all up ports of 2 leafs)
and only 2 ports of the leaf SW\_L1\_2 holding just 2 nodes.
{\scriptsize\begin{verbatim}
| $ curl http://localhost:12345/tenants/1/l1Ports
| [
|     { "pNum": 3, "sName": "SW_L1_2" },
|     { "pNum": 4, "sName": "SW_L1_2" },
|     { "pNum": 1, "sName": "SW_L1_3" },
|     { "pNum": 2, "sName": "SW_L1_3" },
|     { "pNum": 3, "sName": "SW_L1_3" },
|     { "pNum": 4, "sName": "SW_L1_3" },
|     { "pNum": 1, "sName": "SW_L1_4" },
|     { "pNum": 2, "sName": "SW_L1_4" },
|     { "pNum": 3, "sName": "SW_L1_4" },
|     { "pNum": 4, "sName": "SW_L1_4" }
| ]
\end{verbatim}}
\subsubsection{A bad request example:}
Now let's see what happens if we try to over-provision the cluster by requesting a tenant of $13=32-20+1$ hosts:
{\scriptsize\begin{verbatim}
| $ curl http://localhost:12345/tenants -d "id=2" -d "n=13" -X POST
| {
|     "message": "Fail to allocate tenant 2"
| }
\end{verbatim}}

\subsubsection{Delete tenant 1:}
{\scriptsize\begin{verbatim}
| $ curl http://localhost:12345/tenants/1 -X DELETE
\end{verbatim}}
We now have a command file under OSCfg/ that deletes the OpenStack tenant and aggregate
{\scriptsize\begin{verbatim}
| #!/bin/bash
| #
| # Removing tenant 1 from OpenStack
| #
| nova aggregate-remove-host laas-aggr-1 comp-11 >> cmd-3.log
| nova aggregate-remove-host laas-aggr-1 comp-12 >> cmd-3.log
| nova aggregate-remove-host laas-aggr-1 comp-13 >> cmd-3.log
| nova aggregate-remove-host laas-aggr-1 comp-14 >> cmd-3.log
| nova aggregate-remove-host laas-aggr-1 comp-15 >> cmd-3.log
| nova aggregate-remove-host laas-aggr-1 comp-16 >> cmd-3.log
| nova aggregate-remove-host laas-aggr-1 comp-17 >> cmd-3.log
| nova aggregate-remove-host laas-aggr-1 comp-18 >> cmd-3.log
| nova aggregate-remove-host laas-aggr-1 comp-19 >> cmd-3.log
| nova aggregate-remove-host laas-aggr-1 comp-20 >> cmd-3.log
| nova aggregate-delete laas-aggr-1 >> OSCfg/cmd-3.log
| keystone tenant-delete laas-tenant-1 >> OSCfg/cmd-3.log
\end{verbatim}}
\subsubsection{Retry allocating the 13 nodes tenant:}
{\scriptsize\begin{verbatim}
| $ curl http://localhost:12345/tenants \
|    -d "id=2" -d "n=13" -X POST
| {
|     "N": 13,
|     "hosts": 13,
|     "l1Ports": 13,
|     "l2Ports": 0
| }
\end{verbatim}}

\end{enumerate}

\subsection{Running Simulation of LaaS algorithm}
In this section we provide instruction for the simulation of a LaaS engine handling a large number of tenant requests.
The procedure provided here is similar to the one used to obtain the results in the paper.
In the paper, we also used a scheduler that implements the \emph{Simple} and the \emph{Unconstrained} algorithms.

\begin{enumerate}

\item \emph{Choose your cluster topology:}\\
For example the maximal full bisection 3 level XGFT with 36 port switches is:
XGFT(3; 18,18,36; 1,18,18)
It has 11,628 hosts, 648 L1, 648 L2 and 324 L3 switches.

\item \emph{Generate a set of tenant requests:}\\
We do that by running the utility bin/genJobsFlow:
For this example we use an exponential distribution with an average of 8 hosts.
The tenant run time is uniformly distributed in the range [20,3000].
Please try --help to see other possible options.
{\scriptsize\begin{verbatim}
| ./bin/genJobsFlow -n 10000 -s 8 -r 20:3000 -a 0 > \
    examples/exp=8_tenants=1000_arrival=0.csv
\end{verbatim}}

\item \emph{Run the simulator}\\
After we have prepared the tenant requests file and decided about the topology we can run:
{\scriptsize\begin{verbatim}
| $ python ./src/sim.py -m 18,18,36 -w 1,18,18 \
|          -c examples/exp=8_tenants=1000_arrival=0.csv
| -I- Obtained 10000 jobs					
| -I- first waiting job at: 20 lastJobPlacementTime 10623	
| -I- Total potential hosts * time = 1.23673e+08 		
| -I- Total considered jobs: 9976 skip first: 0 last: 24	
| -I- Total actual hosts * time = 1.17281e+08		
| -I- Host Utilization = 94.83 %				
| -I- L1 Up Links Utilization  = 38.36 %			
| -I- L2 Up Links Utilization  = 10.70 %			
| -I- Total Links Utilization  = 48.40 %			
| -I- Run Time = 14.2 sec
\end{verbatim}}
The details of each allocation/deallocation are provided in the log file: isol.log.
Each line describes one transaction and contains the total hosts/links as well as
their detailed indices within the topology.

\item \emph{Check that the results are legal:} \\
The checker needs to know the topology size. So it requires this info on the command line:
{\scriptsize\begin{verbatim}
   checkAllocations -n/--hosts-per-leaf n
                    -k/--num-l1-per-l2 m2
                    -1/--total-l1s t1
                    -2/--total-l2s t2
                    -3/--total-l3s t3
                    -l/--log log-file
\end{verbatim}}
{\scriptsize\begin{verbatim}
| $ ./bin/checkAllocations -n 18 -k 18 -1 648 -2 648 -3 324 \
|        -l isol.log
| -I- Checked 10000 ADD and 8760 REM jobs
| -I- Added/Rem 35573/30117 L1PORTS and 567/482 L2PORTS
\end{verbatim}}

\end{enumerate}

\section{Experimental Setup}\label{sec:experiment}

\subsection{Hardware}
The experiment was run on the 32-node cluster presented in~\fig{fig:test_cluster}.
The hosts are of two types:
\begin{itemize}
\item 30 hosts are HP ProLiant DL320e G8 E3-1220v2 B120i 2x1Gb 1x8GB 1x500GB HOT PLUG DVD-RW 350W 3Y. Each containing 4-core Intel Xeon CPU E3-1220 V2 at 3.10GHz.
\item 2 hosts are IBM System x3450 servers featuring Intel Xeon processors 2.80 GHz and 3.0 GHz/1600 MHz, with 12 MB L2, and 3.4 GHz/1600 MHz, with 6 MB L2.
\end{itemize}

The InfiniBand NICs are: MHQH19-XTC Single 4X IB QDR Port, PCIe Gen2 x8, Tall Bracket, RoHS-R5 HCA Card, QSFP Connector.
The InfiniBand switches are: MIS5024Q-1BFR 36-port non-blocking 40Gb/s unmanaged Switch System.

\subsection{Software}
The machines run Scientific Linux release 6.5 (Carbon).
The MPI used is mvapich2-2.0rc2.
Our experiment uses a simple MPI program that executes an MPI\_AllToAll collectives or 2 dimensional stencil communications using ISend/IRecv followed by MPI\_Barrier.
The programs are provided under the sub-directory \textbf{mpi\_experiment}.
This directory also holds the RUN script that was used to invoke each of the 4 tenants MPI applications with a delay after invoking the previous. The host files used are also included.

\begin{table*}
	\centering
    \begin{tabular}{ | l | l | p{0.7 \columnwidth} | }
    \hline
    Parameter & Value & Description \\ \hline
    MACRelayUnitPP.bufferSize	& 65,536	& Per port buffer size, meaning total buffer size = bufferSize*numRealPorts \\
    MACRelayUnitPP.processingTime &	3.00E-07	& Switch processing delay \\
    TCP.advertisedWindow & 65,535 & Receiver window of TCP \\
    TCP.delayedAcksEnabled	&  false & No delayed ACKS \\
    TCP.minRexmitTimeout & 0.3	& Minimal retransmission timeout \\
    TCP.mss	& 1452	& TCP MSS \\
    TCP.nagleEnabled & true	& TCP parameter \\
    TCP.tcpAlgorithmClass &  DCTCPNewReno	& DCTCP based on NewReno is used \\
    TCPScatterGatterClientApp.idleInterval	&  exponential(200us)	& Time between successive Shuffles (computation time) \\
    TCPScatterGatterClientApp.reconnectInterval	& 1.00E-06	& Time to setup the new connection \\
    TCPScatterGatterClientApp.replyLength	& 2	& Resolver reply just ACK \\
    TCPScatterGatterClientApp.requestLength	& 65,536	& Example data size of 64KiB from Mapper to Resolver \\
    \hline
    \end{tabular}
    \caption{Ethernet model (iNET) parameters and their values}
    \label{tbl:inet_sim_params}
\end{table*}

\section{Ethernet Simulations}\label{sec:eth_sim}
For simulation of an Ethernet-based topology we used an enhanced iNET framework.
We base our code on iNET 2.2 and extend it with DCTCP modules.
The switch forwarding is also enhanced with ECMP-like forwarding with a hash function that works modulo $(SrcHostIndex + DstHostIndex, NumberOfUpPorts)$.
The parameters used by the simulator are described in the following Table~\ref{tbl:inet_sim_params}.

The application used to generate the MapReduce Shuffle stage is an application that runs Scatter and then Gather from a list of nodes. To mimic the Shuffle, the Scatter provides parallel send of the Mapper data size and the Gather is of size 2 bytes only.

The tenants are placed on hosts numbered:\\
Tenant 1:  7,  3, 25, 18,  0, 13, 24, 12\\
Tenant 2:  8,  9, 20, 29,  6,  1, 28,  5\\
Tenant 3: 11,  4, 16, 21,  2, 17, 22, 14\\
Tenant 4: 19, 15, 10, 27, 31, 26, 23, 30

\section{InfiniBand Simulations}\label{sec:ib_sim}
The InfiniBand simulation utilizes Mellanox published model~\cite{zahavi_infinibandtm_2010}.
We have enhanced this model with an application that relies on MPI semantics and is able to replay MPI traces.
The parameters used for our simulation are provided in Table~\ref{tbl:ib_flit_sim_params}.

The tenants that are placed on the 1,728-node cluster are of the sizes:
\begin{itemize}
  \item Two tenants: the two are 810 and 834.
  \item Eight tenants: all are 216 nodes.
  \item Thirty two tenants: all are 54 nodes.
\end{itemize}

The tenants execute cycles of computation and communication. The computation time is of uniform distribution in the range $[5,15]\mu sec$. So the traffic to computation ratio for Stencil application exchanging 32KB of data on each dimension is: Calculation = $10\mu sec$. Communication = $\frac{32KB}{4GB/sec} = 8\mu sec$. So the ratio of ideal computation to communication is 10/8 for 32KB exchanges. 
For all-to-all shuffles we increase the computation time to be uniform in the range $[20,80]\mu sec$, but the data is sent to each other node in the tenant. So for a 32KB exchange on an 8-node tenant, the ratio of computation to ideal communication time is $\frac{50}{7*8} = 50/56$.

\begin{table*}[]
	\centering
    \begin{tabular}{ | l | c | p{1.2 \columnwidth} | }
    \hline
    Module.Parameter & Value & Description \\ \hline
    IBGenerator.flit2FlitGap& 0.001& A gap inserted between flits [nsec] \\
    IBGenerator.flitSize & 64 & The flit size (IB credit is 64 bytes) \\
    IBGenerator.genDlyPerByte& 2.5e-10& Speed of generating bytes [sec/B] \\
    IBGenerator.maxContPkts& 10& Maximum number of continuous packets of same application \\
    IBGenerator.pkt2PktGap& 0.001& Gap inserted between packets [nsec] \\
    IBGenerator.popDlyPerByte& 2e-10& speed of popping up data to next layer [sec/B] \\
    IBInBuf.maxBeingSent& 3& Switch speedup - number of parallel packets being drained from input buffer \\
    IBInBuf.maxStatic0& 800& Buffer size [credits] \\
    IBInBuf.maxVL& 0& Maximal VL simulated \\
    IBInBuf.width& 4& Link withs is 4 lanes \\
    IBOutBuf.credMinTime& 0.256& Maximal time between credit updates [usec] \\
    IBOutBuf.maxVL& 0& Maximal VL simulated \\
    IBOutBuf.size& 66& Host output buffer size [B] \\
    IBOutBuf.size& 78& Switch port output buffer size [B] \\
    IBSink.flitSize& 64& The flit size (IB credit is 64 bytes) \\
    IBSink.hiccupDelay& 1e+06& The receiver may hiccup for 1usec \\
    IBSink.hiccupDuration& 0.0001& Length of a hiccup \\
    IBSink.maxVL& 0& Maximal VL simulated \\
    IBSink.popDlyPerByte& 2.5e-10& Speed of removing Bytes to the PCIe \\
    IBVLArb.busWidth& 24& Input bus width of the switch arbiter \\
    IBVLArb.coreFreq& 250,000,000& Switch core frequency \\
    cModule.ISWDelay& 50& Intrinsic latency of the switch input buffer [nsec] \\
    cModule.VSWDelay& 50& Intrinsic latency of the switch arbiter [nsec] \\
    \hline
    \end{tabular}
    \caption{InfiniBand model parameters and their values}
    \label{tbl:ib_flit_sim_params}
\end{table*}

\end{appendices}

\end{document}